\documentclass[10pt,journal,compsoc]{IEEEtran}

\usepackage[colorlinks,urlcolor=blue,linkcolor=blue,citecolor=blue]{hyperref}
\usepackage[most]{tcolorbox}

\usepackage{tabularx}
\usepackage{listings}
\usepackage{tikz}
 
\usepackage{pgfplots}
\pgfplotsset{compat=1.18}

\usepackage{booktabs}   %% For formal tables:
\usepackage{subcaption} %% For complex figures with subfigures/subcaptions
\usepackage[labelfont=bf]{caption}
\usepackage[utf8]{inputenc} 
\usepackage[T1]{fontenc}
\usepackage{microtype}

\usepackage{enumitem}
\usepackage{color}
\usepackage{amssymb}
\usepackage{multirow}
\usepackage{etoolbox}
\usepackage[plain]{algorithm}% http://ctan.org/pkg/algorithm
\usepackage[noend]{algorithmic}% http://ctan.org/pkg/algorithm
\usepackage{syntax}
\usepackage{comment}
\usepackage{amsthm}
\usepackage{amsmath}
\usepackage{xspace}
\usepackage{tcolorbox}
\usepackage[all]{nowidow}
\usepackage{booktabs}
\usepackage{algorithmic}
\usepackage[english]{babel}
\usepackage{hyphenat}
\usepackage{ulem}
\usepackage{listings}
\usepackage{mdframed}

\definecolor{pblue}{rgb}{0.13,0.13,1}
\definecolor{pgreen}{rgb}{0,0.5,0}
\definecolor{pred}{rgb}{0.9,0,0}
\definecolor{pgrey}{rgb}{0.46,0.45,0.48}
\definecolor{darkblue}{rgb}{0.0, 0.0, 0.55}
\definecolor{light-gray}{gray}{0.9}

\newcommand{\lstbg}[3][0pt]{{\fboxsep#1\colorbox{#2}{\strut #3}}}

\lstdefinelanguage{diff}{
  basicstyle=\ttfamily\small,
  morecomment=[f][\lstbg{pred!20}]-,
  morecomment=[f][\lstbg{pgreen!20}]+,
  morecomment=[f][\textit]{@@},
  morecomment=[f][\textit]{---},
  morecomment=[f][\textit]{+++},
}

\newcommand{\StudyOneFaultySubjects}{2{,}665}
\newcommand{\StudyOneCompilableSubjects}{2{,}335}

\newcommand{\csSmall}{5,000}
\newcommand{\csSmallCE}{2,665}
\newcommand{\csSmallOK}{2,335}

\newcommand{\clang}{\textsc{clang-1600.0.26.6}}
\newcommand{\clangSimp}{\textsc{clang}}

\newcommand{\gpt}{\textsc{ChatGPT-5.2}}
\newcommand{\codex}{\textsc{Codex-GPT5.5}}

\newcommand{\gptoss}{\textsc{GPT-OSS-20B}}
\newcommand{\gptossTime}{55.5h}
\newcommand{\gptossTimeInstance}{39.9s}

\newcommand{\typechef}{\textsc{TypeChef}}
\newcommand{\typechefPrecision}{1.00}
\newcommand{\typechefRecall}{0.24}
\newcommand{\typechefAccuracy}{0.59}
\newcommand{\typechefFScore}{0.39}

\newcommand{\gemini}{\textsc{Gemini 3.6 Flash}}

\newcommand{\geminiTimeInstance}{7.7\,s}   % média por sujeito
\newcommand{\geminiTime}{46.1\,min}        % total das 357 chamadas

\newcommand{\geminiRestored}{182}
\newcommand{\geminiRestoredPct}{95.8\%}
\newcommand{\geminiCandidatesCompile}{349}
\newcommand{\geminiPrecision}{0.90}
\newcommand{\geminiRecall}{1.00}
\newcommand{\geminiAccuracy}{0.94}

\newcommand{\geminiTP}{190}\newcommand{\geminiTN}{147}
\newcommand{\geminiFP}{20}\newcommand{\geminiFN}{0}
\newcommand{\geminiCost}{7 USD}

                                                  {}

\newcommand{\revision}[1]{{\color{black}#1\normalfont}}
\definecolor{green}{RGB}{0,100,0}

\begin{document}

\title{\revision{An Empirical Study of Foundation Models for Variability-Induced Compilation Errors in Configurable C Code}}

\author{
    \IEEEauthorblockN{
        Rohit Gheyi\IEEEauthorrefmark{3}, Lucas Albuquerque\IEEEauthorrefmark{3}, Márcio Ribeiro\IEEEauthorrefmark{2}, Eduardo Almeida\IEEEauthorrefmark{6}, Danyllo Albuquerque\IEEEauthorrefmark{4}, Mirko Perkusich\IEEEauthorrefmark{4} \\
    }
    \IEEEauthorblockA{
        \IEEEauthorrefmark{3}Federal University of Campina Grande (UFCG), Brazil\\Email: rohit@dsc.ufcg.edu.br and lucas.albuquerque@ccc.ufcg.edu.br\\
    }
    \IEEEauthorblockA{
        \IEEEauthorrefmark{2}Federal University of Alagoas, Brazil\\Email: marcio@ic.ufal.br\\
    }
    \IEEEauthorblockA{
        \IEEEauthorrefmark{6}Federal University of Bahia, Brazil\\Email: eduardo.almeida@ufba.br\\          
    }
    \IEEEauthorblockA{
        \IEEEauthorrefmark{4}VIRTUS Research, Development and Innovation Center/UFCG, Brazil\\Email: 
    \{danyllo.albuquerque,mirko\}@virtus.ufcg.edu.br\\
    }    
}

\maketitle

\begin{abstract} In configurable systems, conditional compilation can hide compilation errors under untested feature combinations. We investigate foundation models for detecting such errors and, in a controlled setting, \revision{restoring compilability in configurable C code. Study~I evaluates \gptoss{} on \csSmall{} synthetic snippets generated by \gpt{} from 30 curated seeds and exhaustively compiled under all Boolean feature assignments; it also compares \typechef{} and evaluates \gemini{} on a stratified sample. \gptoss{} achieved 84.7\% micro-precision and 52.1\% micro-recall for affected configurations. Coverage depended on reporting style: presence conditions covered 99.4\% of failing configurations, whereas explicit enumerations covered 29.5\% under a prompt requesting only a minimal justifiable set. \gptoss{} restored compilability for 1,930 of 2,665 faulty snippets (72.4\%), while \gemini{} did so for 182 of 190 sampled faulty snippets (95.8\%). A paired counterfactual audit found no evidence that an identified label-correlated \texttt{\#define} property materially influenced \gptoss{}'s predictions. Study~II evaluates \codex{} on 100 faulty file-level subjects from five mature configurable systems and reports target-fault-aligned problems in 94 subjects, including four of five historical bugs.} Overall, foundation models can support localized detection, explanation, and triage, but should complement compiler-based and variability-aware analyses; compiler acceptance does not establish semantic correctness. \end{abstract}

\section{Introduction}
\label{s:introduction}

In configurable systems, compilation errors may arise only under specific feature combinations and remain hidden during development and testing. Ensuring compilation correctness is therefore difficult, especially as variability annotations evolve in long-lived systems~\cite{liebig-2010}. Prior work has linked variability-related complexity to lower code quality, maintainability, and developer productivity~\cite{Malaquias:2017,DBLP:conf/wcre/BaxterM01}.

Conditional compilation based on \texttt{\#ifdef} remains widely used because of its flexibility and low implementation overhead~\cite{flavio-ecoop2015}. However, traditional compilers analyze one concrete configuration per invocation, so failures in untested combinations may remain hidden. Numerous variability-related bugs have been documented in configurable systems~\cite{abal-2014,Abal18,Medeiros16,flavio-gpce-2013,flavio-gpce2015,flavio-sbes-2020}. Variability-aware parsers such as \typechef{}~\cite{typechef} and \textsc{SuperC}~\cite{superc} reason about multiple configurations simultaneously, but their practical use may require reconstructing include paths, macro definitions, feature constraints, platform assumptions, and other build context.

Large language models (LLMs) have shown strong capabilities in program understanding, code generation, and software review~\cite{Goodfellow-et-al-2016,DBLP:conf/nips/VaswaniSPUJGKP17}, motivating their application to testing, debugging, and program analysis~\cite{se-llms-2023,teste-llms-2024}. Their role in compilation-related tasks remains underexplored~\cite{compiler-llm-hpc2026}. \revision{To the best of our knowledge, no prior systematic empirical study has evaluated whether foundation models can detect variability-induced compilation errors in configurable C code and, in a compiler-based controlled setting, generate transformations that restore compilability.}

\revision{In this article, we investigate foundation models for analyzing variability-induced compilation errors in configurable C code. We examine whether they can detect compilation failures caused by feature variability and, in a controlled setting, generate transformations that restore compilability. In our setting, the models reason about feature macros and conditional-compilation behavior, but they neither systematically explore the complete configuration space nor provide soundness or completeness guarantees.}

\revision{Our evaluation comprises two complementary studies. Study~I constructs a controlled benchmark of \csSmall{} synthetic configurable C snippets generated by \gpt{} from 30 small configurable C examples curated in prior work. We exhaustively compile every Boolean feature assignment to derive the labels and validate candidate transformations. Because the benchmark is model-generated, we also audit its provenance, diagnostic coverage, structural repetition, near-duplicates, and label-correlated properties. Study~II evaluates localized target-fault detection in 100 file-level subjects from BusyBox, Gnuplot, Apache HTTP Server, OpenSSL, and Vim through a target-file-centered IDE workflow with repository context. It includes five historical bugs and 95 mutation-based subjects constructed in the corresponding pre-fix revisions.}

\revision{Study~I shows that configuration coverage depends strongly on reporting style. 
Under a prompting strategy that requested only a minimal confidently justified set, explicit enumerations covered 29.5\% of the affected configurations, whereas reported presence conditions covered 99.4\%; complete configuration sets still require external verification. \gptoss{} restores compilability for 1,930 of 2,665 faulty snippets (72.4\%). In a stratified sample of 357 subjects, \gemini{} detected all 190 faulty subjects but misclassified 20 of the 167 compilable subjects; it restored compilability for 182 faulty subjects (95.8\%). These results establish compiler acceptance, not semantic correctness or behavior preservation. The artifact audit also identifies class-correlated structural regularities, including a conditional \texttt{\#define} property that perfectly separates the binary labels. A paired counterfactual audit does not detect a material effect of this property on \gptoss{}'s predictions in the evaluated sample, but it cannot exclude reliance on other correlated structures or broader synthetic-artifact bias.}

\revision{In Study~II, \codex{} reports a problem aligned with the target fault in 94 of 100 subjects, including four of five historical bugs and 90 of 95 mutation-based subjects. Because all Study~II subjects contain a known target fault, this result measures target-fault detection sensitivity rather than precision or accuracy over faulty and fault-free files. Together, the studies provide compiler-validated controlled evidence from synthetic snippets, limited direct evidence from historical bugs, and controlled detection evidence from injected faults in real project files. The findings indicate that foundation models can support localized detection, explanation, and triage, but should complement rather than replace compiler-based and symbolic variability-aware analyses.}

\revision{In summary, the main contributions are as follows:
\begin{itemize}
\item \textbf{Study~I: controlled evaluation and benchmark audit.} We construct and exhaustively compiler-validate \csSmall{} synthetic configurable C snippets, evaluate subject-level detection, configuration coverage, compilability restoration, and repeated-run stability, and audit the model-generated benchmark for provenance, repetition, diagnostic coverage, and label-correlated properties. We also characterize outputs alongside \typechef{} and evaluate \gemini{} on a stratified sample (Section~\ref{s:study-synthetic-snippets}).
\item \textbf{Study~II: target-file-centered evaluation.} We evaluate \codex{} on 100 file-level subjects containing a documented or injected target fault from five mature configurable C projects through an IDE workflow with repository context (Section~\ref{s:study-real-systems}).
\item \textbf{Replication artifacts.} We release the datasets, prompts, model responses, analysis scripts, and supporting artifacts~\cite{artefatos}, including the deterministic metric pipeline and a script that regenerates the Study~I oracle from source.
\end{itemize}}

Section~\ref{s:motivating-example} presents a motivating example, Section~\ref{s:study-overview} summarizes the empirical design, Sections~\ref{s:study-synthetic-snippets} and~\ref{s:study-real-systems} present the two studies, Section~\ref{s:threats-validity} discusses threats to validity, Section~\ref{s:related} reviews related work, and Section~\ref{s:conclusions} concludes the article.

\section{Motivating Example}
\label{s:motivating-example}

Configurable systems are software systems that support multiple variants derived from a single shared code base. These variants are obtained by enabling or disabling configuration options, allowing developers to tailor software to different platforms, hardware constraints, and functional requirements. In the context of C systems, such variability is typically implemented through preprocessor directives such as \texttt{\#ifdef}, which control the inclusion or exclusion of code fragments at compile time. Due to their simplicity and flexibility, these mechanisms are widely adopted in practice.

Throughout this article, we adopt the terminology of Software Product Lines (SPL) to describe configurable C systems. Specifically, we treat preprocessor macros as features, each valid combination of macro assignments as a product (or configuration), and the overall configurable system as a software product line. This conceptual mapping is well established in the SPL literature and provides a standard abstraction for reasoning about variability in systems that rely on conditional compilation~\cite{ClementsNorthrop2001}.

However, this flexibility comes at a cost. Each configuration option introduces conditional code, and combinations of options may interact in subtle and non-obvious ways. As the number of configuration options grows, the number of possible variants increases exponentially, making it increasingly difficult for developers to fully understand, test, and maintain the system. In particular, errors may arise only under specific combinations of options, remaining hidden during standard development and testing~\cite{flavio-ecoop2015,flavio-tse-2018}. Prior work has shown that the way variability is managed has a direct impact on code quality, maintainability, and system reliability~\cite{Malaquias:2017,flavio-ecoop2015,aldo-emse-2021}.

To illustrate this problem concretely, we present a small but representative example that shows how interactions among a few configuration options can lead to a compilation error affecting only a single variant of the system.

\subsection{Variability-Aware Compilation Error}

\revision{
Consider the simplified code fragment shown in
Listing~\ref{lst:example-busybox}, adapted from a historical issue in
BusyBox, a widely used collection of Unix utilities for embedded
systems. The issue was introduced in commit \texttt{7291755}.
For presentation purposes, we simplify the surrounding code and
represent the relevant configuration options as \texttt{M1} and
\texttt{M2}.}\footnote{\href{https://busybox.net}{BusyBox Project: Commit \texttt{7291755}}}

\begin{lstlisting}[basicstyle=\footnotesize,language=C, label=lst:example-busybox, caption={\revision{Simplified code adapted from a historical BusyBox issue.}}]
#if defined(M1) && defined(M2)
  struct info {
    const char *name;
    const char *pw;
  };
#endif
#if defined(M2)
  struct info userinfo;
#endif
\end{lstlisting}

\revision{
For this simplified example, we assume that \texttt{M1} and
\texttt{M2} are independent Boolean options and consider all four
possible assignments. Here, value \texttt{1} denotes that the macro is defined and value \texttt{0} denotes that it is undefined. In a complete configurable system, a feature model or build system could rule out some combinations.
}
\begin{enumerate}
  \item \textbf{\texttt{M1=0, M2=0}}: no code is included, and the program compiles successfully.
  \item \textbf{\texttt{M1=1, M2=0}}: neither the type nor the variable is present, and compilation also succeeds.
  \item \textbf{\texttt{M1=1, M2=1}}: both the type definition and the variable declaration are included, again resulting in successful compilation.
  \item \textbf{\texttt{M1=0, M2=1}}: \revision{the variable \texttt{userinfo} is declared while the definition of \texttt{struct info} is absent, causing a compilation error.}
\end{enumerate}

\revision{
The error results from inconsistent presence conditions: the declaration
of \texttt{userinfo} is enabled whenever \texttt{M2} is enabled,
whereas the required type definition is available only when both
\texttt{M1} and \texttt{M2} are enabled. Thus, neither option is faulty
in isolation; the failure emerges from their interaction~\cite{SoaresSMA18}.
Such configuration-dependent declaration-use inconsistencies are
particularly difficult to detect when only a subset of variants is
routinely built and tested~\cite{iran-ist-16}.
}

\subsection{Traditional Compilers}

\revision{
Traditional compilers, such as \clangSimp{}, analyze one concrete
configuration per compilation. They correctly report the error in
Listing~\ref{lst:example-busybox} when invoked with \texttt{M1=0}, \texttt{M2=1}, but they do not automatically establish whether another untested assignment fails. Consequently, if the problematic configuration is not explicitly built and tested, the error may remain undetected.
}

\revision{
In systems with many configuration options, such as the Linux kernel,
the configuration space can become too large for exhaustive
enumeration~\cite{linux-survey-2026}. Developers therefore rely on
selected builds, sampling strategies, testing, or specialized analyses,
which may leave rare feature interactions unchecked.
}

\subsection{Variability-Aware Tools}

\revision{
To analyze conditional compilation beyond one concrete configuration,
variability-aware tools such as \typechef{}~\cite{typechef} and
\textsc{SuperC}~\cite{superc} preserve variability during preprocessing
and parsing. This allows them to reason about code belonging to
different configurations without compiling every variant separately.
Their practical application, however, depends on reconstructing relevant
analysis context, including include paths, macro definitions, platform
assumptions, build information, and configuration constraints. This
setup can be nontrivial in large and evolving projects.
}

\revision{
\typechef{} is a well-known variability-aware parsing framework for
configurable C systems. Unlike a standard compilation pipeline that
preprocesses one concrete configuration, \typechef{} preserves
conditional-compilation information and associates source-code elements
with presence conditions---Boolean formulas describing the
configurations in which those elements occur. It can therefore
represent alternative program fragments in a variability-aware syntax
tree and analyze many configurations without explicitly enumerating
all products.
}

\revision{
This design makes \typechef{} an appropriate symbolic baseline for
interactions between C constructs and preprocessor variability. Its
practical coverage nevertheless depends on the available build context,
including include paths, macro definitions, feature constraints, and
platform assumptions. Incomplete context may lead to conservative
diagnostics, infeasible reports, or missed compiler-observable
problems. Our comparison is therefore limited to the evaluated
benchmark and \typechef{} configuration rather than intended as a
general comparison with all variability-aware analyses.
}

\subsection{Foundation Models}

\revision{
Recently, foundation models have emerged as a promising complementary
approach for a wide range of software engineering tasks
~\cite{se-llms-2023,teste-llms-2024}. These models are trained on large
corpora of code and text and exhibit strong capabilities in understanding
code structure, relating declarations and uses, and generating
human-readable explanations and candidate fixes. Unlike traditional
analysis tools, they do not rely on explicitly defined rules for every
configuration scenario.
}

\revision{
Foundation models may therefore be useful for inspecting conditional
code and identifying configuration-dependent inconsistencies with
limited project-specific setup. However, unlike symbolic
variability-aware tools, they do not provide sound or exhaustive
configuration-space guarantees. Their role in variability-induced
compilation-error analysis consequently remains an empirical question.
A recent survey on LLM-enabled compilation highlights the lack of
studies evaluating whether such models can detect and repair compilation
errors caused by configuration variability~\cite{compiler-llm-hpc2026}.
}

\revision{
This example captures the central problem examined in this article:
whether foundation models can recognize compilation failures caused by
inconsistent feature-dependent code. Study~I evaluates this capability
under controlled and exhaustively compiled snippets, whereas Study~II
examines localized detection in substantially larger files from real
configurable projects.
}

\section{\revision{Empirical Study Overview}}
\label{s:study-overview}

Before presenting the studies in detail, we summarize their empirical designs and the roles of the evaluated models and tools.

Study~I provides a controlled evaluation on \csSmall{} synthetic configurable C snippets. Exhaustive compilation of every Boolean feature assignment enables precise subject-level and configuration-level oracles and compiler-based validation of end-to-end compilability restoration. \gptoss{} is the main evaluated model, \gemini{} provides a complementary proprietary-model comparison on a stratified sample, and \typechef{} serves as a representative variability-aware parsing baseline. Study~I also examines reporting style, repeated-run stability, and structural properties of the model-generated benchmark. \gpt{} supports dataset construction and prompt refinement but is not evaluated.

Study~II evaluates \codex{} in a target-file-centered IDE workflow with access to repository context. It contains 100 faulty file-level subjects from five mature configurable C projects: one historical variability-related bug and 19 mutation-based subjects per project, constructed in the corresponding pre-fix revision using patterns inspired by prior variability bugs~\cite{abal-2014,Abal18,flavio-gpce-2013,flavio-gpce2015,Braz:2016,Braz:2018}. Unlike Study~I, Study~II evaluates localized target-fault detection rather than compilability restoration or exhaustive configuration coverage. It therefore emphasizes realistic project context while retaining a predefined target-fault oracle.
Table~\ref{tab:model-task-matrix} summarizes the role, task, and scope of each model and tool.

\begin{table*}[t] \centering \caption{\revision{Role, type, task, and scope of each model and tool used in the evaluation.}} \label{tab:model-task-matrix} \tiny \begin{tabular}{lllll} \toprule \textbf{Model/Tool} & \textbf{Type} & \textbf{Role} & \textbf{Task} & \textbf{Scope} \\ \midrule \gpt{} & Proprietary foundation model & Preparation support; not evaluated & Dataset synthesis and metaprompting & Study~I dataset; prompts for Studies~I--II \\ \gptoss{} & Open-weight foundation model & Main evaluated model in Study~I & Detection and compilability restoration & Synthetic configurable C snippets \\ \gemini{} & Proprietary foundation model & Complementary comparison & Detection and compilability restoration & Stratified sample of Study~I \\ \codex{} & IDE-based coding assistant & Evaluated model in Study~II & Per-file target-fault detection & Faulty C files from real configurable projects \\ \typechef{} 0.3.7 & Variability-aware parser & Baseline tool; not a foundation model & Detection only & Study~I synthetic configurable C snippets \\ \bottomrule \end{tabular} \end{table*}

We selected these models and tools to cover complementary usage scenarios rather than to rank available foundation models or analysis techniques exhaustively. \gptoss{} supports locally controlled and reproducible execution; \gemini{} provides an additional proprietary-model perspective on the Study~I tasks; \codex{} represents an IDE-based coding-assistant workflow with repository access; and \typechef{} represents symbolic reasoning over preprocessor variability. 

We use foundation models as the broadest term for models trained at scale on large and diverse corpora and adaptable to multiple downstream tasks. We use large language models (LLMs) when referring specifically to language-based foundation models or when following the terminology adopted in the broader literature. We use coding models for foundation models specialized, trained, or adapted for code-related tasks, and coding agents for systems that combine a model with tools, an execution environment, and project or repository context. We use proprietary models for closed models accessed through hosted services, in contrast to open-weight models that can be executed locally.

Across both studies, we investigate whether foundation models can support the analysis of variability-induced compilation errors. We use variability-aware in an assistive sense: the evaluated models reason about feature macros and conditional-compilation behavior but neither systematically explore the complete configuration space nor provide soundness or completeness guarantees. The resulting evidence has three distinct forms: (i) compiler-validated controlled evidence from synthetic snippets; (ii) limited direct evidence from five historical bugs; and (iii) controlled target-fault detection evidence from 95 injected faults in real project files. We report these forms separately because they provide different levels of experimental control, and protection against direct answer memorization.

\section{Study I: Synthetic Configurable C Snippets}
\label{s:study-synthetic-snippets}

This section presents our first study, which evaluates whether foundation models can detect compilation errors and restore compilability in synthetic configurable C snippets. 

\subsection{Methodology}
\label{s:methodology}

\begin{figure*}[!htbp]
    \centering
    \includegraphics[width=0.85\textwidth]{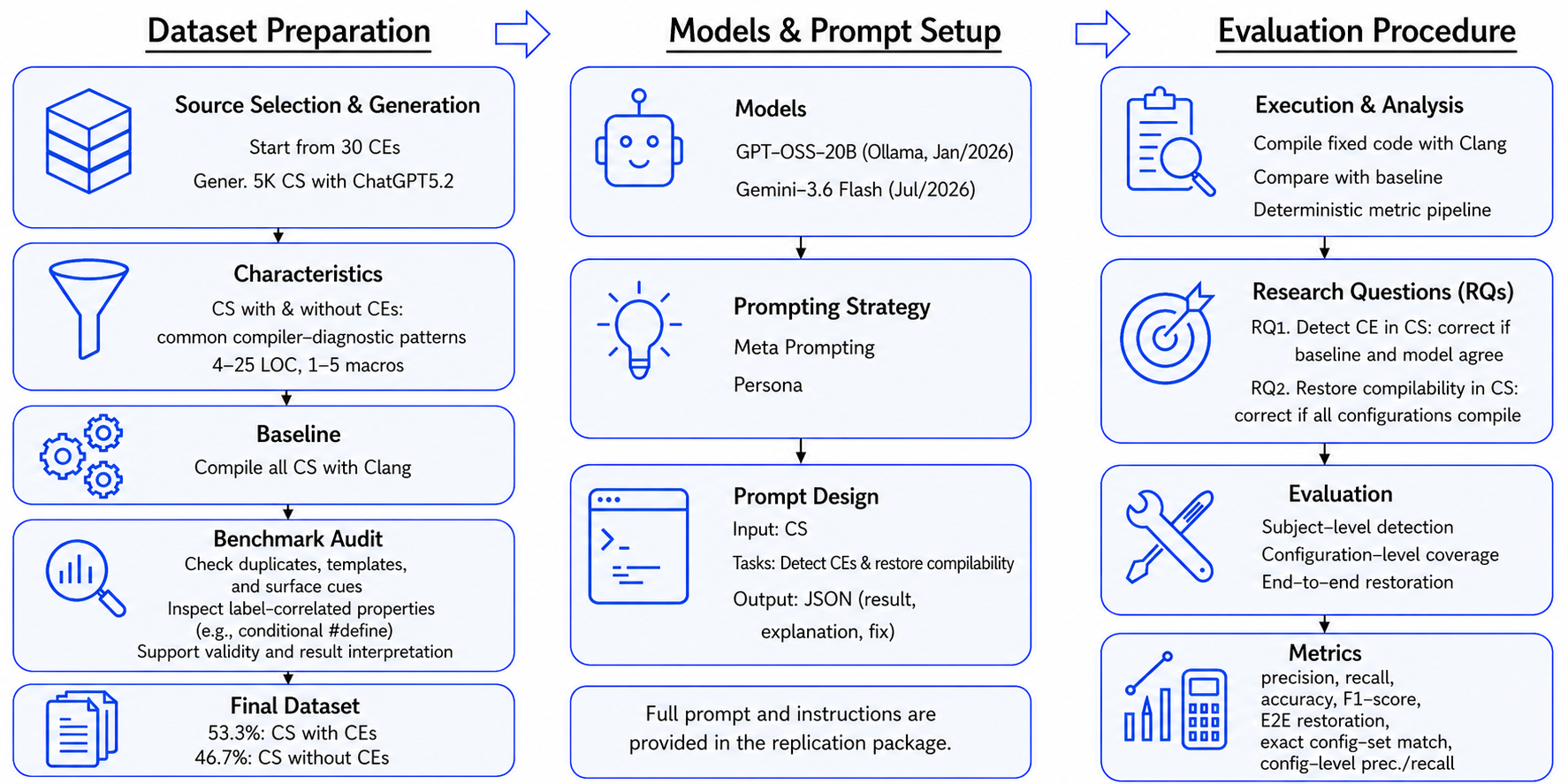}
    \caption{\revision{Methodology for evaluating models to detect compilation errors and restore compilability in synthetic configurable C snippets.}}
    \label{fig:methodology}
\end{figure*}

Figure~\ref{fig:methodology} summarizes the methodology.

\subsubsection{GQM}

The goal~\cite{basili1994thegqm} of this study is formulated as follows: \textit{Analyze foundation models} for the purpose of \textit{evaluating their effectiveness} with respect to \textit{detecting compilation errors and restoring compilability through generated candidate transformations} from the viewpoint of \textit{software developers} in the context of \textit{synthetic configurable C snippets}. Based on this goal, we derive the following research questions (RQ):
\begin{itemize}
\item[\textbf{RQ$_{1}$}]
To what extent can \gptoss{} detect whether a synthetic configurable C snippet contains at least one non-compiling configuration, and what coverage of the affected configurations does it achieve under the evaluated prompting strategy?
\item[\textbf{RQ$_{2}$}] To what extent can \gptoss{} restore compilability for erroneous synthetic configurable C snippets by generating code that compiles successfully across all derived configurations?
\end{itemize}

\revision{We compare model outputs with the compiler-derived oracle at two granularities. Subject-level detection determines whether at least one configuration fails and uses ``contains a non-compiling configuration'' as the positive class. Configuration-level coverage compares the reported and compiler-derived failing sets through exact-set agreement, micro-precision, and micro-recall. For RQ$_2$, end-to-end restoration requires an originally non-compiling subject, a reported error, a nonempty candidate, and successful compilation under every derived configuration. Compiler acceptance does not establish semantic correctness, behavioral equivalence, preservation of intended functionality, or absence of non-compilation defects.}

\subsubsection{Dataset}
\label{sec:methodology-dataset}

\noindent\textit{Creation and baseline.} Prior datasets of variability-related compilation faults are relatively small~\cite{abal-2014,Abal18,Braz:2016,Braz:2018}. We therefore used the 30 small configurable C examples curated in prior work~\cite{DBLP:conf/sbes/AlbuquerqueG024} as seeds in the prompt to \gpt{}, which generated \csSmall{} synthetic snippets targeting both compilable and non-compiling instances. 
Only three of the 30 seed examples contain a \texttt{\#define} directive. All 30 seeds are non-compiling, so the seed collection contains no compilable class at all and therefore no counterpart to the association observed in the generated benchmark, in which the directive co-occurs exclusively with compilable subjects. For the 5,000 generated snippets, the synthesis process did not record seed-to-instance links or transformation traces. An audit of the available metadata, subject ordering, and normalized structural similarity could not reconstruct reliable links to particular seed examples. We therefore cannot report generated-instance counts per seed. The perfect label association reported below therefore cannot have been inherited from the seed collection, although its origin in the synthesis process cannot be traced to particular seeds or transformations.
A post hoc category-level comparison found that seven of the 15 distinct seed error labels had a plausible corresponding normalized first-diagnostic pattern in the generated benchmark, whereas the remaining eight had no such counterpart; this aggregate correspondence does not establish seed-to-instance provenance.
The complete seed catalog, audit criteria, category mapping, and analysis scripts are available in the replication package~\cite{artefatos}.

In January~2026, every Boolean assignment was compiled with \clang{} on macOS Tahoe 26.1 using \texttt{-std=c99}, and only compiler-derived labels were retained. The final benchmark contains \csSmallCE{} subjects with at least one non-compiling configuration; all configurations compile for the remaining subjects. This independent validation is essential for a model-generated benchmark, whose characterization is necessarily post hoc.

\noindent\textit{Characteristics.} The snippets contain 4--25 LOC (median 16; mean 14.7), 1--5 feature macros (median 3), and 0--16 non-compiling configurations (median 1; mean 2.9).
\revision{
We assessed duplication and structural repetition using two normalized representations. After removing comments, blank lines, and whitespace differences, all \csSmall{} snippets remained unique. An abstract representation that replaces identifiers, macro names, user-defined types, and literals while preserving C and preprocessor structure grouped the benchmark into 45 templates containing 46--537 subjects (median 63). Using Python's \texttt{difflib.SequenceMatcher}, a
Ratcliff--Obershelp-style sequence-matching algorithm, over the basic normalization at a conservative threshold of 0.95, we identified 382 highly similar unordered pairs involving 120 subjects (2.4\% of the benchmark). Because mutually similar groups may generate several pairwise matches, these counts are not independent. The benchmark therefore contains no exact duplicates under the basic normalization, but repeatedly instantiates a limited set of structures and includes a small concentration of near-duplicates.

The audit also identified a perfect label association: all \StudyOneCompilableSubjects{} compilable subjects contain at least one conditional \texttt{\#define}, whereas none of the \StudyOneFaultySubjects{} non-compiling subjects does. All 2,644 occurrences appear inside conditional regions, so the property is a legitimate code construct rather than an explicit label marker, but a trivial rule based on it still achieves 100\% subject-level accuracy. The 45 templates are also class-pure---36 occur only among non-compiling subjects and nine only among compilable subjects---but this observation is not independent because the abstract representation preserves preprocessor syntax. We therefore treat both findings jointly as class-correlated structural regularities and do not interpret binary metrics alone as evidence of variability-aware reasoning. Section~\ref{sec:counterfactual} tests whether changing the \texttt{\#define} property materially affects \gptoss{}'s predictions.

Among the 2,665 non-compiling subjects, the first compiler error is a declaration/type-specifier error in 1,291 cases (48.4\%), an identifier/name-binding error in 1,071 (40.2\%), and an expression/operator error in 303 (11.4\%). Among the compilable subjects, 309 produce only macro-redefinition warnings and 2,026 produce no compiler diagnostic. The snippets combine conditional directives with common C constructs such as structures, global variables, and occasional function declarations; they cover a controlled set of declaration, type, name, and expression failures rather than the broader range of C build failures.

Table~\ref{tab:subject-characterization} summarizes their variability characteristics. The benchmark contains only disciplined conditional blocks, nesting depth one, single-macro conditions, and no multi-branch alternatives. Thus, Study~I exercises small, mostly sequential declaration--use inconsistencies rather than the full interaction space illustrated in Section~\ref{s:motivating-example}. Study~II partially addresses this limitation with files reaching nesting depth six and containing undisciplined annotations.
}

\begin{table}[t]
\centering
\caption{\revision{Characterization of the benchmark subjects.}}
\label{tab:subject-characterization}
\scriptsize
\begin{tabular}{lrrr}
\toprule
\textbf{Metric} & \textbf{Min.} & \textbf{Median} & \textbf{Max.} \\
\midrule
LOC & 4 & 16 & 25 \\
Number of macros & 1 & 3 & 5 \\
Number of configurations & 2 & 8 & 32 \\
Conditional regions & 1 & 4 & 7 \\
Max. conditional nesting depth & 1 & 1 & 1 \\
Undisciplined conditional blocks & 0 & 0 & 0 \\
Multi-macro conditions & 0 & 0 & 0 \\
Multi-branch conditional regions & 0 & 0 & 0 \\
\bottomrule
\end{tabular}
\end{table}

\subsubsection{Prompting Strategy}
\label{s:methodology-prompt}

We employ Meta Prompting~\cite{hou-metaprompting,prompts,prompt-techniques} to guide the model in generating or refining prompts. Specifically, we asked \gpt{} the following query:

\begin{tcolorbox}[
  breakable,
  colback=gray!5,
  colframe=black!60,
  boxrule=0.5pt,
  arc=2pt,
  left=6pt,right=6pt,top=6pt,bottom=6pt
]
\footnotesize
Create a prompt that receives C code and checks whether there is a compilation error in any configuration. If there is, present the corrected code. Do not use \texttt{define} or \texttt{undefine} macros. The code must be corrected. The JSON output must be structured as follows: 
\begin{itemize}
    \item Result: indicating the configurations that do not compile, separated by \texttt{;}. If there are none, return an empty string.
    \item Explanation: a brief explanation of the compilation error.
    \item Corrected code: fix the code. If it is not possible to determine the fix, leave it blank.
\end{itemize}
\end{tcolorbox}

In response, the model generated a prompt design specifically tailored to identifying and fixing compilation errors in \revision{synthetic configurable C snippets}, as detailed next.

\begin{tcolorbox}[
  breakable,
  colback=gray!5,
  colframe=black!60,
  boxrule=0.5pt,
  arc=2pt,
  left=6pt,right=6pt,top=6pt,bottom=6pt
]
\footnotesize
You are an expert C engineer specialized in configurable systems (C code with \#if/\#ifdef feature flags). \\
 \\
\noindent LANGUAGE STANDARD \\
\noindent Assume the code must be valid according to the ISO C99 standard.
Use C99 rules for syntax, declarations, and semantics when reasoning about compilation errors. \\
 \\
\noindent INPUT \\
\noindent You will receive a single C source file written in C (ISO C99). The code is provided below. \\
 \\
\noindent <code> \texttt{code} </code> \\
 \\
\noindent TASK \\
\noindent 1) Consider the software product line induced by the feature macros used in the file. A ``product'' is any assignment of those macros to enabled/disabled (on/off). \\
\noindent 2) Determine whether there exists at least one product configuration for which the code does NOT compile under a standard C99 compiler (e.g., gcc or clang with -std=c99). \\
\noindent 3) If at least one product fails to compile: \\
\noindent - Identify the failing products as boolean assignments over macros (e.g., ENABLE_PAM=1,ENABLE_AUTH_MD5=0). \\
\noindent - List only products that fail. If there are many, list the minimal set of distinct failing products you can confidently justify from the code. \\
\noindent 4) Provide a corrected version of the C code that compiles for ALL products under C99, by fixing the code itself. \\
\noindent - You MUST NOT rely on external build flags, compiler options, or macro redefinitions (-D/-U). \\
\noindent - You MUST NOT remove variability or delete features unless absolutely necessary. \\
\noindent - Prefer localized fixes: align declarations and uses under the same conditions, add missing declarations, introduce safe defaults, refactor conditional code, etc. \\
\noindent - Preserve the intended behavior as much as possible. \\

\noindent OUTPUT FORMAT (MANDATORY — JSON ONLY) \\
\noindent Return a single valid JSON object with exactly the following keys: \\ 
\noindent "result": "<semicolon-separated list of failing products, or empty string if none>",  \\
\noindent "explanation": "<short explanation of the compilation issue(s), or a brief statement that no failing products exist>",  \\
\noindent "fixed_code": "<full corrected C source code (C99-compliant), or empty string if you cannot confidently fix it>" \\

\noindent CONSTRAINTS \\
\noindent - If no failing products exist, set "result" to an empty string (""). \\
\noindent - If you cannot confidently produce a correct fix, set "fixed_code" to an empty string (""). \\
\noindent - Do NOT include markdown, comments outside the JSON, or any additional keys. \\
\noindent - The output must be strictly valid JSON and directly machine-readable.
\end{tcolorbox}

\revision{Here, \texttt{code} is the input snippet. The prompt requested only a minimal confidently justified set of failing products rather than exhaustive configuration enumeration. Preliminary executions showed that omitting the language standard led to inconsistent assumptions across C versions, while weak repair constraints encouraged external macro assignments or removal of variability. We therefore fixed ISO C99, prohibited reliance on external \texttt{-D}/\texttt{-U} flags, requested localized changes, and required machine-readable JSON. These instructions do not guarantee preservation of the original macro vocabulary, as discussed in Section~\ref{s:results-rq2}.}

\subsubsection{Models}
\label{sec:studyone-models}

We selected \gptoss{} as an open-weight model supporting reproducible local execution. We used Ollama through the LangChain API~\cite{ollama-setup} on a Mac Mini M4 Pro with 64\,GB RAM, retaining the default configuration except for temperature 0.4. The full-dataset and five-run evaluations were conducted in January~2026.

\subsubsection{Evaluation Procedure}
\label{sec:evaluation-procedure}

\revision{
\noindent\textit{Compiler oracle and evaluation criteria.} After running \gptoss{}, we compare its outputs with the \clang{} compiler oracle.
All checks use Apple \clang{} with ISO C99 explicitly enabled through \texttt{clang -std=c99 -c <source-file> <macro-flags>}, where \texttt{<macro-flags>} contains the corresponding \texttt{-D} and \texttt{-U} assignments. A configuration is classified as non-compiling only when the compiler returns a nonzero exit code; warnings are recorded separately, which is important for macro redefinitions.}

\revision{At the subject level, a detection is correct when the model and oracle agree on whether at least one configuration fails. At the configuration level, we compare the predicted and compiler-derived failing sets through exact-set agreement, micro-precision, and micro-recall. A complete assignment denotes one product, whereas a partial assignment is interpreted as a presence condition and expanded over all completions of its unconstrained macros. Responses containing tokens that are not macro-value pairs are excluded. We report both the aggregate over interpretable responses and the values obtained when partial assignments are excluded. The oracle covers preprocessing, parsing, and type-checking failures reported during compilation, but not warnings, link-time errors, or build-system failures.}

\revision{For RQ$_2$, a subject is restored end-to-end only when it originally contains a non-compiling configuration, the model reports an error, supplies a nonempty candidate, and the candidate compiles under every derived configuration using the same invocation. Empty candidates count as unsuccessful.}
\revision{Because Study~I contains no feature models or explicit feature constraints, every Boolean assignment over the macros in a snippet is treated as a valid product. This choice may include combinations that a complete configurable system would exclude, but it enables exhaustive and reproducible validation of the controlled benchmark.}
\revision{A deterministic pipeline computed all Study~I metrics from the compiler-based oracle and recorded compiler invocations, hashes, column mappings, outputs, and a reproducibility manifest. Warning-only diagnostics were treated as successful compilations, and candidate transformations were validated by recompiling every derived configuration.}

\revision{
\noindent\textit{Reproducibility pipeline and stratified sample.}
We drew a stratified random sample of 357 subjects for the repeated-run stability analysis of \gptoss{} and the complementary evaluation of \gemini{}. The sample contains 167 compilable and 190 faulty subjects, totaling 1,048 failing configurations. Exhaustive compilation also identified 146 warning-only configurations across 20 subjects. The sampling procedure, compiler-derived oracle, and outputs are available in the replication package~\cite{artefatos}.}

\revision{
\noindent\textit{Counterfactual audit of the \texttt{\#define} association.} To distinguish the existence of a label-correlated property from actual model reliance, we conducted a paired counterfactual audit following prior work on shortcut learning and counterfactual tests~\cite{geirhos2020shortcut,elazar2024measuring}. We selected 200 subjects: 100 non-compiling subjects distributed across the three diagnostic categories, macro counts, and all 36 faulty templates, and 100 compilable subjects without diagnostics distributed by macro count. The 309 warning-only subjects were excluded from the removal condition because removing their macro definition would also remove the construct producing the warning.

Each subject contributes an original snippet, a cue variant that changes the lexical presence of \texttt{\#define}, and a local sham variant that applies a comparable edit while preserving the directive's presence or absence. For non-compiling subjects, the cue variant adds a guarded macro definition, whereas the sham variant adds guarded static declarations. For compilable subjects, the cue variant removes the guarded definition and inlines its value, whereas the sham variant renames the macro and changes its value while retaining the directive. These transformations manipulate the cue while controlling for sensitivity to a comparable local edit and are not intended to preserve semantics beyond the compiler oracle.

We exhaustively compiled all original, cue, and sham variants and retained only subjects whose exact failing sets remained unchanged. Each variant was then evaluated twice with \gptoss{} in a deterministically randomized order, totaling 1,200 calls. We compare paired prediction changes, changes in the direction predicted by the structural rule, changes under the sham condition, and test--retest variation using exact McNemar tests and one-sided 95\% Clopper--Pearson upper bounds.
}

\subsection{Results}
\label{s:results}

\subsubsection{RQ$_{1}$. Detecting Compilation Errors in Snippets}
\label{s:results-rq1}

\revision{
\noindent\textit{Subject-Level Detection.} \gptoss{} produced 2,597 true positives, 2,109 true negatives, 226 false positives, and 68 false negatives. This corresponds to 92.0\% precision, 97.4\% recall, 94.1\% accuracy, 90.3\% specificity, and an F$_1$-score of 0.946. These values describe high observed subject-level performance on the controlled benchmark: the model missed relatively few non-compiling subjects, while most false positives are concentrated in warning-only macro-redefinition cases, discussed next.
The perfect association reported in
Section~\ref{sec:methodology-dataset} limits the interpretation of these binary metrics: the presence of \texttt{\#define} alone perfectly separates the benchmark labels. This association does not, however, establish that \gptoss{} relied on this property. We therefore complement the subject-level results with within-class, configuration-level, and counterfactual analyses.
}

Listing~\ref{lst:id30} presents a \revision{synthetic configurable C snippet} with a single feature macro (\texttt{A}) that conditionally introduces a macro redefinition of \texttt{B}.

\begin{lstlisting}[basicstyle=\footnotesize,language=C, label=lst:id30, caption={Warning-only macro redefinition correctly treated as compilable by \gptoss{}.}]
#if defined(A)
  #define B 1
  #define B 2
#endif
int main() { 
  return 0; 
}
\end{lstlisting}

In the first field of its JSON output, \gptoss{} correctly returns an empty value, indicating that it considers the \revision{snippet} to compile successfully. In the explanation field, the model identifies the duplicated macro definition but classifies it as a warning rather than a compilation error, noting that it does not prevent successful compilation, as shown next.

\begin{mdframed}[backgroundcolor=gray!7, linecolor=black, linewidth=0.5pt]
\footnotesize
\noindent The only issue is that when \texttt{A} is defined, the macro \texttt{B} is defined twice with different replacement lists, which violates a constraint and typically results in a warning but does not prevent compilation. \ldots
\end{mdframed}
\clang{} reports a similar warning. Moreover, the model output proposes a compiler-accepted fix that removes the macro redefinition, even though the code already correctly compiles.

As another example, Listing~\ref{lst:id178} shows a \revision{synthetic configurable C snippet} with two feature macros (i.e., \texttt{A} and \texttt{B}). Under our experimental setup, \clang{} reports a compilation error because the declaration of \texttt{p} violates the C99 standard: implicit \texttt{int} types are not supported in ISO~C99 and later. Specifically, the statement \texttt{p = "str";} appears as a standalone expression after the declaration \texttt{int *p = 0;}, which leads to an invalid top-level construct. In contrast, \gptoss{} does not report any compilation error for this configuration.

\begin{lstlisting}[basicstyle=\footnotesize,language=C, label=lst:id178, caption={\gptoss{} does not \revision{report a} compilation error.}]
#if defined(A)
  int x = 2;
#endif
#if defined(B)
  int *p = 0; p = "str";
#endif
int main(){
  return 0;
}
\end{lstlisting}

As a complementary case, Listing~\ref{lst:id38} presents a \revision{synthetic configurable C snippet} with four feature macros (\texttt{A}, \texttt{B}, \texttt{C}, and \texttt{D}), in which macro \texttt{D} is defined twice when feature \texttt{B} is enabled. According to \clang{} under our experimental setup, this duplication triggers only a warning indicating that \texttt{D} is redefined, but does not prevent successful compilation. In contrast, \gptoss{} incorrectly classifies this scenario as a compilation error.

\begin{lstlisting}[basicstyle=\footnotesize,language=C, label=lst:id38, caption={\gptoss{} \revision{reports a} compilation error.}]
#if defined(A)
  int x = 6;
#endif
#if defined(C)
  int y = 6;
#endif
#if defined(B)
  #define D 1
  #define D 2
#endif
int main() {
  return 0;
}
\end{lstlisting}

\sloppy
A closer inspection reveals that \gptoss{} struggles primarily with low-level syntactic and declaration-related compilation errors. Several missed cases correspond to classic grammar violations, such as \texttt{expected expression} and \texttt{expected ';' after top-level declarator}, as well as C99-specific diagnostics where implicit typing is disallowed (e.g., \texttt{type specifier missing, defaults to 'int'}). These errors often require precise token-level reasoning and strict adherence to the C grammar, which appears challenging for the model. In addition, \gptoss{} failed to detect multiple declaration consistency errors, including redefinitions with incompatible types (e.g., \texttt{redefinition of 'p' with a different type}) and conflicting function or variable types (e.g., \texttt{conflicting types for \textit{cfrrymj}}). Finally, some missed cases involve unresolved name references, such as \texttt{use of undeclared identifier}, which typically arise from subtle mismatches in the visibility of declarations across conditional compilation blocks.

\revision{
For each subject, the pipeline selects the first \texttt{error} or \texttt{fatal error} observed in the deterministic configuration order and classifies its message using ordered regular-expression rules. If no error exists, the first warning is used only for characterization. This procedure yields mutually exclusive first-observed diagnostic categories rather than an inventory of every diagnostic produced across all configurations. The three compiler-error categories contain the 2,665 non-compiling subjects. The remaining 2,335 subjects compile successfully: 2,026 produce no diagnostic and 309 produce only macro-redefinition warnings.
}

\begin{table}[t]
\centering
\caption{\revision{Dataset distribution, within-category
subject-level detection accuracy, and end-to-end (E2E) compilability restoration by diagnostic category. Because each diagnostic category is label-homogeneous, detection
accuracy corresponds to recall for compiler-error categories and specificity for compilable categories. Restoration is reported only for subjects with compiler errors.}}
\label{tab:error-categories}
%\footnotesize
\scriptsize
\begin{tabular}{lrrcc}
\toprule
\textbf{Diagnostic category} &
\textbf{N} &
\textbf{\%} &
\textbf{Detect.} &
\textbf{\revision{E2E restor.}} \\
\midrule
Declaration/type-specifier       & 1291 & 25.8\% & 95.2\% & \revision{57.9\%} \\
Expression/operator              &  303 &  6.1\% & 99.0\% & \revision{93.1\%} \\
Identifier/name-binding          & 1071 & 21.4\% & 99.7\% & \revision{84.1\%} \\
Warn.-only macro redefin. & 309 & 6.2\% & 29.8\% & \revision{--} \\
No compiler diagnostic           & 2026 & 40.5\% & 99.6\% & \revision{--} \\
\bottomrule
\end{tabular}
\end{table}

\revision{
Detection accuracy was high for all three compiler-error categories. In contrast, accuracy was only 29.8\% for the 309 warning-only macro-redefinition subjects: \gptoss{} correctly treated 92 as compilable but incorrectly classified 217 as containing a compilation error. Thus, the lower accuracy in this category reflects difficulty distinguishing warnings from errors rather than difficulty detecting an actual preprocessor compilation failure. End-to-end restoration requires both correct fault detection and generated code that compiles under our oracle; it is therefore reported only for subjects containing compiler errors.
}

\revision{
The observational results already show that the model is not equivalent to the trivial \texttt{\#define} classifier. Both the 2,026 no-diagnostic subjects and the 309 warning-only macro-redefinition subjects belong to the same structurally defined compilable class, for which that rule would achieve 100\% accuracy, yet \gptoss{} achieves 99.6\% and 29.8\%, respectively. The contrast shows that the model's outputs depend on code properties beyond the mere presence of \texttt{\#define}; because the poorer result occurs in the warning-only subset, it also shows that the model is substantially worse than the trivial rule for that subgroup. This observational comparison does not determine whether the cue has a smaller causal influence, which is addressed by the counterfactual audit. Section~\ref{sec:counterfactual} reports the direct counterfactual test of this association.
}

\revision{
We further stratified subject-level detection by the number of macros and conditional regions. Table~\ref{tab:performance-variability} reports the number of subjects, the number containing at least one compiler error, recall over faulty subjects, and specificity over compilable subjects. Recall remains high and relatively stable across every group containing faulty subjects, ranging from 96.1\% to 98.5\%. Thus, the number of macros or conditional regions does not show a clear association with the model's ability to recognize faulty subjects in this benchmark. The variation in aggregate accuracy previously observed across groups is primarily explained by class composition and specificity. In the groups containing one, two, or three conditional regions, every compilable subject belongs to the warning-only macro-redefinition category, for which the model frequently reports a compilation error. In contrast, the groups containing six or seven conditional regions contain no faulty subjects, so their high aggregate accuracy reflects specificity only. We therefore avoid interpreting higher aggregate accuracy in these groups as evidence that subjects with more conditional regions are easier to analyze.
}

\begin{table}[t]
\centering
\caption{\revision{Subject-level detection performance by
selected variability characteristics.}}
\label{tab:performance-variability}
\scriptsize
\begin{tabular}{llrrrr}
\toprule
\textbf{Dimension} & \textbf{Value} & \textbf{N} &
\textbf{Faulty N} & \textbf{Recall} &
\textbf{Specificity} \\
\midrule
\multirow{5}{*}{Macros}
& 1 & 495  & 429 & 97.7\% & 22.7\% \\
& 2 & 1,480 & 891 & 98.4\% & 92.9\% \\
& 3 & 1,050 & 489 & 97.1\% & 90.4\% \\
& 4 & 1,021 & 432 & 96.1\% & 93.2\% \\
& 5 & 954  & 424 & 96.9\% & 92.6\% \\
\midrule
\multirow{7}{*}{Conditional regions}
& 1 & 495  & 429 & 97.7\% & 22.7\% \\
& 2 & 469  & 417 & 96.6\% & 23.1\% \\
& 3 & 1,033 & 963 & 98.5\% & 27.1\% \\
& 4 & 1,034 & 432 & 96.1\% & 93.2\% \\
& 5 & 971  & 424 & 96.9\% & 92.9\% \\
& 6 & 524  & 0   & --    & 99.8\% \\
& 7 & 474  & 0   & --    & 99.4\% \\
\bottomrule
\end{tabular}
\end{table}

\textit{Configuration-Level Coverage.} Subject-level detection asks whether at least
one configuration fails, whereas the model output also reports which products are
affected. \revision{The prompt asks for the minimal set of distinct failing products
that the model can confidently justify, and the responses use two distinct reporting
styles. In 2,128 subjects the model enumerates complete concrete macro assignments; in
693 it reports partial assignments, that is, presence conditions that constrain a
subset of the macros and denote the set of products obtained by completing the
remaining ones; in 2,177 it reports no failing product; and 2 contain invalid product
tokens. Scoring only complete assignments would discard the partial ones as
uninterpretable, which measures compliance with an enumeration format rather than the
ability to characterize the affected configuration space. Because a partial assignment
is a valid answer to the prompt as written, and denotes a well-defined set of products,
we score it over that set and report both interpretations.}

\revision{Table~\ref{t:config-coverage} decomposes the results by reporting style. The
two styles behave very differently. When the model enumerates concrete assignments, its
reports are highly precise but cover a minority of the affected products: 93.5\%
micro-precision and 29.5\% micro-recall, with an exact match to the compiler-derived
set in 48.4\% of those subjects. When it reports a presence condition, the denoted set
covers 4,819 of the 4,850 failing configurations of those subjects (99.4\% micro-recall)
at 80.5\% micro-precision, matches the compiler-derived set exactly in 78.8\% of them,
and covers every failing configuration in 686 of 693. Aggregating all interpretable
responses yields 84.7\% micro-precision and 52.1\% micro-recall, against 93.5\% and
28.1\% when partial assignments are excluded. Exact set agreement is almost unchanged
between the two interpretations, at 72.9\% and 73.7\%.}

\revision{Two conclusions follow. First, the aggregate configuration-level recall is
governed by reporting style rather than by an inability to identify affected products:
93.2\% of the residual false negatives arise in the enumerating subgroup, 6.3\% from the
68 subjects the model classified as compilable, and 0.4\% from presence conditions.
Second, incomplete enumeration is a real limitation of the enumerating style and the
conservative prompt instruction, not evidence that the model cannot characterize the
affected configuration space. The \texttt{\#define} shortcut contains no information
about which macro assignments fail, so both interpretations show that the reported
configurations carry information beyond that property. Under the evaluated setting,
compiler-based enumeration or symbolic variability-aware analysis remains necessary when
a complete and verified configuration set is required, and a prompt explicitly demanding
exhaustive enumeration would be needed to estimate coverage under that instruction.}

\begin{table}[t]
\centering
\caption{Configuration-level reporting by response style. Presence conditions are
partial macro assignments, scored over the set of products they denote.}
\label{t:config-coverage}
\scriptsize
\begin{tabular}{lrrrrr}
\toprule
\textbf{Response style} & \textbf{N} & \textbf{Fail. cfg.} & \textbf{Micro-prec.} & \textbf{Micro-rec.} & \textbf{Exact} \\
\midrule
Complete enum. & 2,128 & 9,119 & 93.5\% & 29.5\% & 48.4\% \\
Presence condition   &   693 & 4,850 & 80.5\% & 99.4\% & 78.8\% \\
No failing product   & 2,177 &   438 & --     &  0.0\% & 96.9\% \\
\midrule
All interpretable    & 4,998 & 14,407 & 84.7\% & 52.1\% & 73.7\% \\
Complete only        & 4,305 &  9,557 & 93.5\% & 28.1\% & 72.9\% \\
\bottomrule
\end{tabular}
\end{table}

\subsubsection{RQ$_{2}$. \revision{Restoring Compilability}}
\label{s:results-rq2}

To answer RQ$_2$, we evaluate whether \gptoss{} generates candidate
transformations that satisfy the compiler-based restoration criterion
defined in Section~\ref{sec:evaluation-procedure}.
\revision{
\gptoss{} restored compilability end-to-end for 1,930 of the 2,665 originally non-compiling subjects (72.4\%). This metric requires correct subject-level detection, a nonempty candidate fix, and successful compilation of the generated code under every derived configuration. Separately, the model reported an error and provided a nonempty candidate fix for 2,821 subjects; 2,153 of these candidates compiled in all configurations (76.3\%). The latter value measures generated-code compilability and includes cases in which the original subject already compiled, whereas the 72.4\% result is the RQ$_2$ end-to-end restoration rate. Neither metric establishes semantic correctness or behavior preservation.
}

\revision{
Among the 735 faulty subjects not restored end-to-end, 68 were missed during subject-level detection and 667 received a candidate that still failed in at least one configuration; none failed because a detected faulty subject lacked a candidate. The 667 non-compiling candidates comprise 482 declaration/type-specifier, 167 identifier/name-binding, and 18 expression/operator subjects. Thus, candidate generation and compiler acceptance, rather than initial detection, account for most restoration failures.
}
A similar limitation has been reported in prior work that used smaller open-weight models to generate and analyze single configurations~\cite{debora-icmla-2025}.

Consider the \revision{synthetic configurable C snippet} shown in Listing~\ref{lst:id20}, which contains a single feature macro (\texttt{A}).

\begin{lstlisting}[basicstyle=\footnotesize,language=C, label=lst:id20, caption={Undefined variable detected by \gptoss{}.}]
#if defined(A)
  int x = abc;
#endif
int main() { 
  return 0; 
}
\end{lstlisting}

\gptoss{} correctly identifies the compilation error and provides a clear explanation stating that the variable \texttt{abc} is not declared when the macro \texttt{A} is enabled. However, for this example, \gptoss{} proposes an incorrect fix, shown in Listing~\ref{lst:id20-incorrect}. When compiling this fix with \clang{} on macOS, it reports an error indicating that the initializer element is not a compile-time constant when assigning \texttt{x} to \texttt{abc}. While this behavior may vary across compilers, in our evaluation this fix was classified as incorrect.

\begin{lstlisting}[basicstyle=\footnotesize,language=C, label=lst:id20-incorrect, caption={Incorrect fix proposed by \gptoss{}.}]
#if defined(A)
  int abc = 0;
  int x = abc;
#endif
int main() { 
  return 0; 
}
\end{lstlisting}

\revision{
Compilation errors can admit multiple compiler-accepted transformations~\cite{Ahmed2019TEGCER}. To provide a reproducible structural characterization, we compared the conditional-compilation macro names in the original and generated code. Of the 1,930 end-to-end restored subjects, 1,185 (61.4\%) preserve the exact macro set. The remaining 745 change it: 742 differ by one macro and 3 by two macros. This analysis does not establish preservation of presence conditions, feature behavior, intended functionality, or semantics; transformations that alter the macro vocabulary should therefore be interpreted cautiously.
}

\revision{
End-to-end compilability restoration also varies by compiler-error category (Table~\ref{tab:error-categories}). It is lowest for declaration/type-specifier errors (57.9\%), higher for identifier/name-binding errors (84.1\%), and highest for expression/operator errors (93.1\%). These rates are calculated only over non-compiling subjects and require correct detection, a nonempty candidate fix, and generated code that compiles in all configurations. The pattern indicates that detecting an error and producing a compiler-accepted transformation require different capabilities, particularly when declarations or type information must be introduced or modified.
}

\subsection{Discussion}
\label{s:discussion}

\subsubsection{\revision{Counterfactual Audit of the \texttt{\#define} Association}}
\label{sec:counterfactual}

\revision{Section~\ref{sec:methodology-dataset} identified a perfectly label-correlated structural property: every compilable Study~I subject contains a conditional \texttt{\#define}, whereas no non-compiling subject does. This establishes that the benchmark admits a trivial subject-level rule, but it does not establish that \gptoss{} used that property. To test this directly, the procedure described in Section~\ref{sec:evaluation-procedure} evaluated 200 subjects, each in original, cue, and local sham variants. In July~2026, we conducted a counterfactual audit using the same local model identifier, prompt, and temperature settings as in the main Study~I evaluation. The cue variant changed the lexical presence of \texttt{\#define}, whereas the sham variant applied a comparable local edit while preserving the directive's presence or absence. All 200 cue and 200 sham variants preserved the exact compiler-derived set of non-compiling configurations, and every variant was evaluated twice, producing 1,200 model calls.}

\revision{Table~\ref{tab:counterfactual-audit} summarizes the paired subject-level results. In the first run, the cue manipulation changed 8 of 200 predictions (4.0\%), exactly the same number as the sham condition; in the second, it changed 5 predictions (2.5\%), compared with 4 (2.0\%) under the sham condition. Repeated evaluation of identical inputs produced pooled test--retest change rates of 3.0--3.5\% across the original, cue, and sham variants. Thus, manipulating \texttt{\#define} produced a change rate comparable to both the local-edit control and the model's observed run-to-run variability. Exact McNemar tests comparing original and cue predictions yielded \(p=0.727\) and \(p=1.000\) in the two runs. Changes in the direction predicted by the structural rule were equally frequent under the sham control, which does not alter the directive: 3 and 2 under the cue condition compared with 4 and 1 under the sham condition.}

\revision{Among the 200 subjects, only 3 in the first run and 2 in the second exhibited prediction changes in the direction predicted by the structural rule, corresponding to one-sided 95\% Clopper--Pearson upper bounds of 3.8\% and 3.1\% on the proportion of subjects whose predictions follow that rule. None of these changes occurred for the same subject in both runs. In the compilable removal condition, removing \texttt{\#define} produced no rule-direction change in either run: all 100 cue variants were classified as compilable in both runs. The manipulation inverts the structural rule by construction, since the rule classifies every original variant correctly and every cue variant incorrectly; subject-level accuracy of \gptoss{} nevertheless did not decrease, changing from 97.5\% to 98.5\% in the first run and from 98.0\% to 98.5\% in the second. Because the audited sample contains only non-compiling subjects and compilable subjects without compiler diagnostics, this accuracy is not directly comparable to the 94.1\% of Section~\ref{s:results-rq1}, which includes the 309 warning-only subjects with 29.8\% accuracy. The corresponding Study~I values for the audited strata are 97.4\% recall and 99.6\% specificity, which the audit reproduces closely. Of the 1,200 responses, 1,198 were strictly valid JSON objects. An automated explanation filter flagged one response, but manual inspection by two authors showed that it referred generically to combinations of feature-macro assignments rather than using the presence or absence of a \texttt{\#define} directive as a decision rule.}

\begin{table}[t]
\centering
\caption{\revision{Paired subject-level results of the counterfactual
\texttt{\#define} audit. ``Changes'' denotes prediction flips relative
to the original. ``Rule-direction changes'' denotes flips toward the
prediction of the structural rule; the sham condition provides the
corresponding local-edit control.}}
\label{tab:counterfactual-audit}
\scriptsize
\begin{tabular}{llrrrr}
\toprule
\textbf{Run} & \textbf{Condition} & \textbf{Orig. acc.} &
\textbf{Variant acc.} & \textbf{Changes} &
\textbf{Rule-dir. chg.} \\
\midrule
1 & Cue  & 97.5\% & 98.5\% & 8 & 3 \\
1 & Sham & 97.5\% & 97.5\% & 8 & 4 \\
2 & Cue  & 98.0\% & 98.5\% & 5 & 2 \\
2 & Sham & 98.0\% & 99.0\% & 4 & 1 \\
\bottomrule
\end{tabular}
\end{table}

\revision{We therefore find no evidence that this specific structural property materially drove \gptoss{}'s subject-level predictions in the evaluated sample. The perfect association remains a limitation of the generated benchmark because it makes the labels predictable without reproducing the compiler analysis, so subject-level accuracy alone should not be treated as evidence of variability-aware reasoning or as a general ranking of detector capabilities. However, the counterfactual results do not support the stronger causal interpretation that the observed predictions were driven by this cue. The audit cannot exclude smaller effects, use outside the sampled subjects, or reliance on other correlated properties, and it was conducted only for \gptoss{}.}

\revision{The configuration-level and restoration results provide distinct evidence that the cue itself cannot supply. The presence of \texttt{\#define} does not identify which macro assignments fail and cannot generate a compiler-accepted transformation. Accordingly, the configuration-level micro-precision of 84.7\% over all interpretable responses, and of 93.5\% over complete enumerations, shows that the reported configurations carry information the cue cannot supply. The 52.1\% aggregate micro-recall shows that this information is incomplete overall, although the incompleteness is concentrated in the enumerating style rather than distributed across responses. Similarly, 667 of the 735 end-to-end restoration failures occur after detection because the generated candidate still fails to compile, indicating that candidate generation and compiler acceptance are the dominant bottleneck. These findings do not establish semantic reasoning, repair correctness, or immunity to other benchmark artifacts.}

\subsubsection{Accuracy and Stability}
\label{s:discussion-stability}

\revision{
Foundation models are nondeterministic, and repeated executions on the same input may produce different predictions or candidate fixes. We therefore evaluate both effectiveness and stability across five executions of \gptoss{} on the same sample. Detection and restoration use different success criteria: detection requires agreement with the subject-level compiler oracle, whereas end-to-end restoration additionally requires a nonempty candidate that compiles under every derived configuration.
}

\noindent\textit{\revision{Metrics.}}
\revision{Table~\ref{tab:metrics} defines the repeated-run metrics. Pass@\(k\) is retrospective and oracle-dependent: it shows that at least one successful response exists among the first \(k\) attempts but does not select it operationally.
Tar@\(k\) measures total agreement, cons@\(k\) combines agreement with effectiveness, and success-rate spread@\(k\) measures variation among per-run rates. Even-\(k\) ties contribute 0 to cons@\(k\) because no unique mode exists.}

\begin{table}[t]
\centering
\caption{\revision{Effectiveness and stability metrics used in the repeated-run analysis.}}
\label{tab:metrics}
\scriptsize
\setlength{\tabcolsep}{2pt}
\renewcommand{\arraystretch}{0.95}
\begin{tabularx}{\columnwidth}{
  @{}
  >{\raggedright\arraybackslash}p{2.15cm}
  @{\hspace{0.35em}}
  >{\raggedright\arraybackslash}X
  @{}
}
\toprule
\textbf{Metric} & \textbf{Definition} \\
\midrule

Run Success Rate
& Proportion of subjects successfully handled in one execution.
Detection requires agreement with the subject-level oracle;
restoration additionally requires a nonempty candidate that compiles
under every derived configuration. \\

Mean Success@\(k\)
& Average binary success over all subjects and the first \(k\)
executions. \\

\textit{pass@\(k\)}
& Proportion of subjects with at least one successful outcome among
the first \(k\) executions. \\

\midrule

Success-Rate Spread@\(k\)
& Difference between the highest and lowest success rates among the
first \(k\) executions; lower values indicate less run-to-run
variation. \\

\textit{tar@\(k\)}
& Proportion of subjects whose first-\(k\) binary outcomes are
identical, regardless of correctness. \\

\textit{cons@\(k\)}
& Proportion of subjects with a unique modal outcome that is correct
for detection or successful for restoration; an even-\(k\) tie
contributes 0. \\

\bottomrule
\end{tabularx}
\end{table}

\noindent\textit{\revision{Sampling Strategy.}} To limit computational cost, we used the standard proportion-based sample-size formula with finite-population correction, assuming a 95\% confidence level (\(z=1.96\)), a conservative proportion \(p=0.5\), and a margin of error \(e=0.05\):
\[
n_0 = \frac{z^2\,p(1-p)}{e^2}, \qquad n = \frac{n_0}{1 + \frac{n_0 - 1}{N}}.
\]

\revision{
This calculation yields a target sample size of approximately \(n=357\). We use the same stratified random sample of 357 subjects for the repeated-run analysis of \gptoss{} and the complementary evaluation of \gemini{}. Under the final deterministic compiler oracle, 167 subjects have all configurations compile and 190 contain at least one non-compiling configuration. Thus, 46.8\% of the sample is compilable and 53.2\% is non-compiling, closely matching the corresponding 46.7\% and 53.3\% proportions in the complete benchmark. All five executions use this same fixed sample.
}

\noindent\textit{\revision{Results: Detecting Compilation Errors.}}
\revision{
We executed \gptoss{} five times on the same 357-subject sample. Running the five executions required slightly more than 20 hours. The per-run detection accuracies were 0.938, 0.950, 0.944, 0.941, and 0.961. Thus, per-run accuracy ranges from 0.938 to 0.961, with a maximum spread of 0.023. Mean accuracy over the first \(k\) executions remains between 0.938 and 0.947.
}

\revision{
As shown in Figure~\ref{fig:det-metrics-k}, the model achieves pass@1 of 0.938, pass@2 of 0.972, pass@3 of 0.975, and pass@5 of 0.986. Thus, 352 of the 357 sampled subjects receive at least one correct binary classification within five attempts. This result is retrospective and oracle-based: without an aggregation rule or external validator, a user does not know which response should be selected when attempts disagree.
}

\revision{
Total agreement decreases gradually as additional attempts are included: tar@\(k\) decreases from 1.000 at \(k=1\) to 0.894 at \(k=5\). The cons@\(k\) values are 0.938, 0.916, 0.958, 0.941, and 0.955 for \(k=1,\ldots,5\), respectively. The lower value at \(k=2\) results from 20 subjects for which the two binary predictions disagree and therefore produce no unique mode. The results show that repeated attempts increase the probability that a correct classification is present, but they do not guarantee success for every subject or provide an operational selection rule.
}

\begin{figure}[t]
\centering
\begin{tikzpicture}
\begin{axis}[
    width=0.5\linewidth,
    height=4cm,
    xmin=1, xmax=5,
    ymin=0.88, ymax=1.01,
    xtick={1,2,3,4,5},
    ytick={0.90,0.95,1.00},
    xlabel={$k$},
    ylabel={Score},
    legend style={at={(1.03,0.3)}, anchor=south west, draw=none, fill=none},
    grid=both,
]
\addplot+[mark=*] coordinates {
    (1,0.9383753501)
    (2,0.9719887955)
    (3,0.9747899160)
    (4,0.9747899160)
    (5,0.9859943978)
};
\addlegendentry{pass@k}

\addplot+[mark=square*] coordinates {
    (1,1.0000000000)
    (2,0.9439775910)
    (3,0.9243697479)
    (4,0.9159663866)
    (5,0.8935574230)
};
\addlegendentry{tar@k}

\addplot+[mark=triangle*] coordinates {
    (1,0.9383753501)
    (2,0.9159663866)
    (3,0.9579831933)
    (4,0.9411764706)
    (5,0.9551820728)
};
\addlegendentry{cons@k}

\addplot+[mark=diamond*] coordinates {
    (1,0.9383753501)
    (2,0.9439775910)
    (3,0.9439775910)
    (4,0.9432773110)
    (5,0.9467787115)
};
\addlegendentry{Mean accuracy@k}
\end{axis}
\end{tikzpicture}
\caption{\revision{Accuracy and stability metrics for subject-level compilation-error detection by \gptoss{} across five executions on the 357-subject sample.}}
\label{fig:det-metrics-k}
\end{figure}

\noindent\textit{\revision{Results: End-to-End Compilability Restoration.}}
\revision{
We evaluate restoration stability over the 190 subjects in the sample that contain at least one non-compiling configuration according to the final deterministic oracle. For each attempt, restoration is successful only when the model reports a compilation error, provides a nonempty candidate fix, and the generated code compiles under all derived configurations. Empty candidates are treated as unsuccessful attempts, even when the stored compiler-log field is empty.
}

\revision{
The per-run end-to-end restoration rates are 0.726, 0.705, 0.753, 0.689, and 0.737. The highest and lowest values differ by 0.064, showing greater run-to-run variation than observed for detection. Mean restoration over the first \(k\) executions remains between 0.716 and 0.728.
As shown in Figure~\ref{fig:fix-metrics-k}, pass@\(k\) increases from 0.726 at \(k=1\) to 0.821 at \(k=2\), 0.889 at \(k=3\), and 0.905 at \(k=5\). Thus, 172 of the 190 faulty sampled subjects receive at least one end-to-end compiler-accepted restoration within five attempts. This does not mean that an automated pipeline can select the successful candidate without compiling and validating the alternatives.
Restoration outcomes are substantially less stable than detection outcomes. Tar@\(k\) decreases from 1.000 at \(k=1\) to 0.579 at \(k=5\). Cons@\(k\) equals 0.726, 0.611, 0.732, 0.689, and 0.737 for \(k=1,\ldots,5\), respectively. The lower values at even \(k\) partly reflect binary ties: 40 subjects produce tied outcomes at \(k=2\), and 15 at \(k=4\).
These findings show that repeated querying increases the chance of obtaining at least one compiler-accepted candidate, but end-to-end restoration is less stable than detection. Compiler-based and, where feasible, testing-based validation therefore remains necessary for every generated candidate.
}

\begin{figure}[t]
\centering
\begin{tikzpicture}
\begin{axis}[
    width=0.5\linewidth,
    height=4cm,
    xmin=1, xmax=5,
    ymin=0.50, ymax=1.01,
    xtick={1,2,3,4,5},
    ytick={0.50,0.60,0.70,0.80,0.90,1.00},
    xlabel={$k$},
    ylabel={Score},
    legend style={at={(1.03,0.3)}, anchor=south west, draw=none, fill=none},
    grid=both,
]
\addplot+[mark=*] coordinates {
    (1,0.7263157895)
    (2,0.8210526316)
    (3,0.8894736842)
    (4,0.8947368421)
    (5,0.9052631579)
};
\addlegendentry{pass@k}

\addplot+[mark=square*] coordinates {
    (1,1.0000000000)
    (2,0.7894736842)
    (3,0.6736842105)
    (4,0.6263157895)
    (5,0.5789473684)
};
\addlegendentry{tar@k}

\addplot+[mark=triangle*] coordinates {
    (1,0.7263157895)
    (2,0.6105263158)
    (3,0.7315789474)
    (4,0.6894736842)
    (5,0.7368421053)
};
\addlegendentry{cons@k}

\addplot+[mark=diamond*] coordinates {
    (1,0.7263157895)
    (2,0.7157894737)
    (3,0.7280701754)
    (4,0.7184210526)
    (5,0.7221052632)
};
\addlegendentry{Mean E2E@k}
\end{axis}
\end{tikzpicture}
\caption{\revision{End-to-end compilability-restoration and stability metrics for \gptoss{} across five executions on the 190 originally non-compiling subjects in the sample.}}
\label{fig:fix-metrics-k}
\end{figure}

\subsubsection{\gemini{}}
\label{s:gemini}

In addition to evaluating the open-weight model \gptoss{}, we assess a proprietary foundation model to examine whether the findings extend across model families. Because the \csSmall{} \revision{synthetic configurable C snippets} were synthesized using \gpt{}, we selected a model from a different provider to reduce the threat of a model evaluating artifacts that it originally generated. \revision{To control evaluation costs, we conducted the analysis in July~2026 on the same 357-subject stratified sample used for the repeated-run analysis of Section~\ref{s:discussion-stability}.}

\revision{We executed \gemini{} through its public API with provider-side structured output enabled, setting the response MIME type to \texttt{application/json} and supplying an explicit schema for the three requested keys. We used temperature 0, a one-second inter-request delay, and per-subject checkpointing. The prompt and compiler-derived oracle are those of Sections~\ref{s:methodology-prompt} and~\ref{sec:evaluation-procedure}. Execution required \geminiTime{}. All 357 responses were strictly valid JSON objects containing exactly the requested keys, terminated with finish reason \textsc{stop}, and required no post hoc recovery.}

\revision{At the subject level, \gemini{} produced \geminiTP{} true positives, \geminiTN{} true negatives, \geminiFP{} false positives, and \geminiFN{} false negatives, corresponding to precision \geminiPrecision{}, recall \geminiRecall{}, and accuracy \geminiAccuracy{}. All 20 false positives were warning-only macro-redefinition subjects, whereas the model correctly classified all 147 subjects without compiler diagnostics. Its only observed detection failure mode in this sample was therefore misclassifying macro-redefinition warnings as compilation errors.}

\revision{We also evaluated compilability restoration under the criterion of Section~\ref{sec:evaluation-procedure}. All 190 faulty subjects received a nonempty candidate, and recompilation under every derived configuration showed that \gemini{} restored compilability for \geminiRestored{} subjects (\geminiRestoredPct{}). Overall, \geminiCandidatesCompile{} of the 357 generated candidates compiled under every derived configuration. In all eight unsuccessful cases, the number of non-compiling configurations remained unchanged, indicating that the candidate did not address the responsible construct. One such case introduced an \texttt{extern} declaration for a missing identifier but retained the original non-constant-initializer failure.}

\revision{The comparison with \gptoss{} should be interpreted cautiously because \gemini{} was evaluated in a single execution, whereas \gptoss{} was evaluated five times. On the same 20 warning-only subjects, \gptoss{} correctly classified 3, 9, 5, 4, and 11 subjects across its five executions, corresponding to per-run accuracies of 15.0\%, 45.0\%, 25.0\%, 20.0\%, and 55.0\% (32.0\% across the 100 subject-run predictions), whereas \gemini{} classified none correctly in its single execution. Thus, both models struggled with the warning--error distinction, although \gptoss{} exhibited substantial run-to-run variation. On the same 190 faulty subjects, \gptoss{} achieved per-run restoration rates between 0.689 and 0.753 and a \textit{pass@5} value of 0.905, whereas \gemini{} restored 95.8\% in one execution. However, \gemini{} was evaluated only once and on 357 of the \csSmall{} subjects, so its run-to-run variability was not estimated. We therefore treat these results as complementary evidence across model families rather than as a model ranking.}

\subsubsection{Execution Time}

\revision{The full \gptoss{} execution required \gptossTime{}, the five-run sample slightly more than 20 hours, and the 1,200-call counterfactual audit approximately 10.4 hours (31.3 seconds per call). The average full-dataset time was \gptossTimeInstance{} per \gptoss{} subject. The \gemini{} evaluation of the 357-subject sample required \geminiTime{}, averaging \geminiTimeInstance{} per subject, at a total cost of approximately \geminiCost{}. These costs remain relevant even for small snippets.}

\subsubsection{\typechef{}}
\label{s:discussion-typechef}

\revision{We evaluate \typechef{} as a detection-only symbolic baseline in Study~I. The goal is not to claim that foundation models replace variability-aware parsing, but to contrast a foundation model with a traditional symbolic approach under the same controlled synthetic benchmark. \typechef{} is an appropriate representative baseline because it is public, reproducible, applicable to all Study~I subjects, directly addresses C preprocessor variability, and represents a paradigm distinct from foundation-model prompting. Because it does not synthesize repairs, the comparison is restricted to compilation-error detection.}

\revision{We used the official \typechef{} 0.3.7 JAR distribution~\cite{typechefsite} with OpenJDK Temurin 21.0.7+6 on macOS Tahoe 26.1. For each snippet, we generated a temporary source file and analyzed it using \typechef{} with its standard configuration. We supplied a small local include directory with stubs for \texttt{printf} and \texttt{scanf} through the \texttt{-I} option to avoid diagnostics caused only by missing standard-library declarations. A run was considered successful when variability-aware lexing and parsing completed without failure or exception messages; otherwise, it was treated as a positive detection for the binary comparison. This operational mapping enables comparison with the compiler-based oracle, but it does not imply that the evaluated \typechef{} invocation checks every class of C99 compilation error captured by that oracle.}

\revision{In this setup, \typechef{} achieved a precision of \typechefPrecision{}, recall of \typechefRecall{}, accuracy of \typechefAccuracy{}, and F1-score of \typechefFScore{}. It produced 638 true positives, 2,335 true negatives, 2,027 false negatives, and no false positives. Thus, \typechef{} was highly precise when it reported a problem, but missed many snippets that contained compilation errors according to the compiler-based oracle. This result is consistent with the scope of the evaluated setup: \typechef{} focuses on variability-aware lexing and parsing, whereas the oracle also captures broader C99 compiler diagnostics, including declaration, type-specifier, name-binding, and expression-level errors.}

\revision{Both approaches produced correct subject-level results in 2,744 cases (54.9\%). \typechef{} was exclusively correct in 229 cases (4.58\%), \gptoss{} alone in 1,962 cases (39.2\%), and neither in 65 cases (1.3\%). Given the benchmark's trivial label separability and the different diagnostic scopes of \gptoss{} and \typechef{}, these results should be interpreted descriptively rather than as a general ranking of variability-analysis capabilities. The pattern of agreement and disagreement nevertheless motivates complementary workflows: symbolic tools can provide configuration evidence and diagnostics, while foundation models can help explain findings and generate candidates that are subsequently validated by compilers or variability-aware tools. The comparison is limited to the evaluated \typechef{} version, lightweight parser configuration, benchmark, and operational detection criterion; we do not claim that \typechef{} is the unique or universally best baseline.}

\section{Study~II: Target-File-Centered Detection in Real Configurable Projects}
\label{s:study-real-systems}

This section presents our second study, which evaluates whether \codex{} can detect target compilation faults in configurable C files in 100 file-level subjects from five mature configurable C projects using an IDE-based workflow with repository context. The subjects comprise five documented historical faults and 95 mutation-based faults.

\subsection{Methodology}
\label{s:methodology-real}

\begin{figure*}[!htbp]
    \centering
    \includegraphics[width=0.85\textwidth]{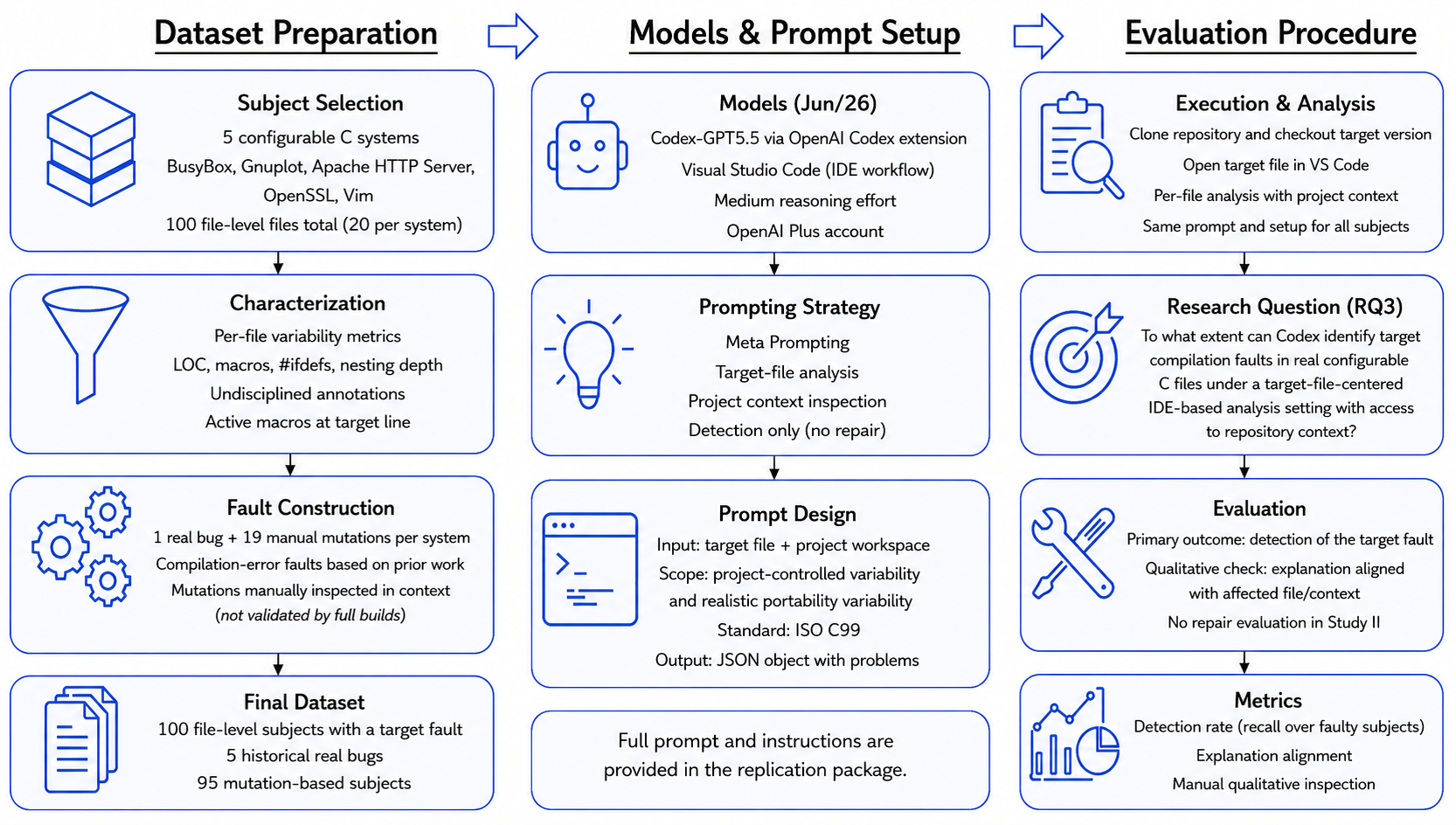}
    \caption{Methodology for evaluating target-file-centered detection of compilation faults in real configurable projects.}
    \label{fig:methodology-real}
\end{figure*}

Figure~\ref{fig:methodology-real} summarizes the Study~II methodology.

\subsubsection{GQM}
\label{s:study-real-systems-gqm}

The goal of this study is formulated as follows: \textit{Analyze an IDE-based foundation-model assistant} for the purpose of \textit{evaluating its effectiveness} with respect to \textit{detecting target compilation faults in individual project files} from the viewpoint of \textit{software developers} in the context of \textit{historical bugs and controlled mutation-based faults in files from real configurable C projects}.

Based on this goal, we derive the following research question:

\begin{itemize}
\item[\textbf{RQ$_{3}$}] To what extent can \codex{} identify target compilation faults in real configurable C files under a
target-file-centered IDE-based analysis setting with access
to repository context?
\end{itemize}

To answer this research question, we combine five historical variability-related compilation bugs with 95 controlled mutation-based subjects. Each subject therefore contains either a documented historical fault or a deliberately introduced compilation-fault pattern. Because Study~II contains no fault-free control subjects, it does not constitute a complete binary-classification evaluation. Our primary quantitative measure is the subject-level detection rate, computed as the proportion of subjects for which the model reports at least one problem aligned with the target fault. This measure is equivalent to recall over the evaluated faulty subjects. Study~II does not estimate precision, specificity, false-positive rate, F$_1$-score, or overall classification accuracy in a population containing both faulty and fault-free files.

\subsubsection{Dataset}
\label{sec:methodology-dataset-real}

We constructed a benchmark containing 100 file-level subjects from five mature configurable projects: BusyBox, Gnuplot, Apache HTTP Server, OpenSSL, and Vim. The benchmark contains 20 subjects per project. For each project, all subjects were derived from the same historical pre-fix revision, i.e., the revision immediately preceding a commit that fixed a documented variability-related compilation bug. The original faulty file constitutes the historical subject, while the other 19 subjects are additional C files from that revision into which we manually introduced one target compilation-fault pattern.

Each file is treated as one subject. This design complements Study~I by exposing the model to substantially larger source files, real project-specific declarations and portability mechanisms, multiple preprocessor annotations, nested conditional-compilation regions, and unrelated surrounding code. However, the use of real project files does not mean that all 100 faults occurred naturally: only five are historical bugs, while the remaining 95 are controlled mutations.

Table~\ref{tab:subjects-characterization} summarizes the main characteristics of the subjects. For each project, we report per-file minima and maxima for file size, distinct macros used in conditional compilation, conditional-compilation annotations, maximum nesting depth, undisciplined annotations, and active enclosing macros at the affected line. Disciplined annotations preserve syntactic structure, whereas undisciplined annotations occur inside syntactic constructs or divide syntactic units.

\begin{table*}[t]
\centering
\caption{Characterization of the 100 Study~II subjects drawn from real configurable projects. Values report per-file minima and maxima for each project.}
\label{tab:subjects-characterization}
\scriptsize
\resizebox{\textwidth}{!}{
\begin{tabular}{lrrrrrrrrrrrrr}
\toprule
\multirow{2}{*}{\textbf{Project}} &
\multirow{2}{*}{\textbf{Subjects}} &
\multicolumn{2}{c}{\textbf{LOC}} &
\multicolumn{2}{c}{\textbf{Macros}} &
\multicolumn{2}{c}{\textbf{Cond. annotations}} &
\multicolumn{2}{c}{\textbf{Depth}} &
\multicolumn{2}{c}{\textbf{Undisciplined}} &
\multicolumn{2}{c}{\textbf{Target macros}} \\
\cmidrule(lr){3-4}
\cmidrule(lr){5-6}
\cmidrule(lr){7-8}
\cmidrule(lr){9-10}
\cmidrule(lr){11-12}
\cmidrule(lr){13-14}
&
&
\textbf{Min} & \textbf{Max} &
\textbf{Min} & \textbf{Max} &
\textbf{Min} & \textbf{Max} &
\textbf{Min} & \textbf{Max} &
\textbf{Min} & \textbf{Max} &
\textbf{Min} & \textbf{Max} \\
\midrule
BusyBox & 20 & 131 & 2,200 & 2 & 21 & 1 & 48 & 1 & 3 & 0 & 2 & 0 & 3 \\
Gnuplot & 20 & 256 & 7,238 & 1 & 50 & 3 & 190 & 1 & 6 & 0 & 6 & 1 & 3 \\
Apache HTTP Server & 20 & 168 & 3,691 & 1 & 33 & 1 & 32 & 1 & 3 & 0 & 2 & 1 & 1 \\
OpenSSL & 20 & 42 & 1,253 & 1 & 34 & 1 & 64 & 1 & 5 & 0 & 0 & 1 & 2 \\
Vim & 20 & 348 & 6,248 & 1 & 43 & 1 & 230 & 1 & 5 & 0 & 2 & 1 & 4 \\
\midrule
\textbf{Total} & \textbf{100} & \textbf{42} & \textbf{7,238} & \textbf{1} & \textbf{50} & \textbf{1} & \textbf{230} & \textbf{1} & \textbf{6} & \textbf{0} & \textbf{6} & \textbf{0} & \textbf{4} \\
\bottomrule
\end{tabular}
}
\end{table*}

Overall, the benchmark contains 130,550 LOC and 95,449 non-blank, non-comment LOC. The selected files range from 42 to 7,238 LOC and contain 988 macro definitions, 1,017 per-file unique macro uses, and 2,630 conditional-compilation annotations. The maximum observed nesting depth is six. The benchmark includes 2,609 disciplined and 21 undisciplined annotations.
The Target macros columns describe the local variability context of the affected line by counting the active enclosing macro conditions. This measure complements the file-level characterization because a file may contain many macros and conditional regions while the affected line is controlled by only a subset of them. It characterizes the local conditional context; it does not by itself establish the complete set of feasible project configurations in which the fault manifests.

\subsubsection{Mutation Procedure and Taxonomy}
\label{sec:methodology-mutations-real}

We use mutation-based subjects as controlled fault-seeding scenarios, not as substitutes for naturally occurring bugs. This design follows the mutation-testing literature, which commonly uses mutants as proxies for faults and has investigated their relationship with real faults~\cite{mutation-testing,Just2014MutantsRealFaults,Papadakis2018MutationScores,Petrovic2021MutationTesting}. Because this relationship is imperfect and context-dependent, we interpret the 95 mutations as controlled detection scenarios in real project files, whereas the five historical bugs provide limited direct evidence on naturally occurring variability-related faults.
The historical subjects were selected from commits previously reported in the variability literature~\cite{Abal18,flavio-gpce-2013,flavio-gpce2015,Braz:2018,Medeiros16}. For each case, we checked out the immediately preceding revision, in which the documented bug was still present. The mutation-based subjects were then created in 19 C files from that same revision.

The injected mutations were inspired by fault patterns reported in prior studies and by the historical bugs inspected during dataset construction. They include conditional-compilation inconsistencies, feature-dependent declaration and use mismatches, name-binding faults, function-interface inconsistencies, invalid pointer or member access, and local syntactic faults. The faulty variants did not occur in the original repositories, which reduces the possibility of direct memorization of the modified code and its corresponding answer. However, the surrounding project code and similar fault patterns may still have appeared in model training data.

We manually inspected each modified file to verify that the edit instantiated its intended fault pattern and preserved the surrounding project and conditional-compilation context. We did not, however, rebuild every project under all relevant feature and portability configurations to reproduce all 95 injected failures through complete project builds. The mutation-based oracle therefore records the deliberately modified element and its intended compilation-fault pattern rather than constituting an exhaustive build-validated oracle over the complete configuration space.

We also recorded the local variability context of each affected line. Ninety-nine of the 100 affected lines occur within code controlled by at least one conditional-compilation macro: 82 subjects have one active enclosing macro, 12 have two, four have three, and one has four. The only exception is the historical BusyBox bug, whose affected line lies outside a conditional-compilation region. These counts characterize the local placement of the target faults, but they do not establish that every macro assignment corresponds to a feasible project configuration.

Each historical bug or mutation was assigned one primary category according to the affected program element and the principal reasoning required to identify the fault. Local syntax faults were classified as lexical/punctuation or delimiter/block-structure faults; consistency faults involving declarations, uses, calls, fields, or types as identifier/name-binding, function interface/call, declaration/type-specifier, or pointer/member/array-access faults; and faults directly affecting directives or feature conditions as preprocessor/feature-condition faults. As shown in Table~\ref{tab:study2-results-category}, the most frequent mutation categories concern pointer/member/array access, function interfaces and calls, and identifier/name binding, which often require relating the edited construct to declarations, types, or uses elsewhere in the file or translation unit. The benchmark also includes local syntax, control-flow, initialization, and preprocessor faults. The five naturally occurring subjects were retained as a separate historical-bug provenance group because their number was too small for a meaningful category-level breakdown.

\begin{table}[t]
\centering
\caption{Study~II subject distribution and detection results by fault category or provenance.}
\label{tab:study2-results-category}
\scriptsize
\setlength{\tabcolsep}{4pt}
\renewcommand{\arraystretch}{0.95}
\begin{tabular}{lrrr}
\toprule
\textbf{Fault category or provenance}
& \textbf{Sub.}
& \textbf{Detected}
& \textbf{Rate} \\
\midrule
Pointer/member/array access mutation & 19 & 16 & 84.2\% \\
Function interface/call mutation & 17 & 17 & 100.0\% \\
Identifier/name-binding mutation & 15 & 15 & 100.0\% \\
Delimiter/block-structure mutation & 12 & 11 & 91.7\% \\
Declaration/type-specifier mutation & 8 & 8 & 100.0\% \\
Expression/operator mutation & 7 & 7 & 100.0\% \\
Lexical/punctuation mutation & 7 & 7 & 100.0\% \\
Historical real-world variability bug & 5 & 4 & 80.0\% \\
Return-value/control-transfer mutation & 4 & 3 & 75.0\% \\
Control-structure mutation & 3 & 3 & 100.0\% \\
Initialization/literal mutation & 2 & 2 & 100.0\% \\
Preprocessor/feature-condition mutation & 1 & 1 & 100.0\% \\
\midrule
\textbf{Total} & \textbf{100} & \textbf{94} & \textbf{94.0\%} \\
\bottomrule
\end{tabular}
\end{table}

\subsubsection{Prompting Strategy}
\label{s:methodology-prompt-real}

As in Study~I, we used Meta Prompting~\cite{hou-metaprompting,prompts,prompt-techniques} to refine the model instructions. However, the Study~II prompt differs in two important ways. First, it focuses only on detection, not repair, because validating repairs in real configurable systems would require reproducible builds, project-specific dependencies, generated configuration headers, platform assumptions, and representative feature configurations. Second, it approximates an IDE-based workflow: \codex{} was instructed to analyze a target file in the currently opened VS Code workspace and to inspect other project files only when needed to understand that file or its translation unit.

The prompt constrained the analysis to ISO C99 compilation errors and instructed the model not to edit files, propose patches, or report warnings, runtime bugs, memory bugs, link-time errors, or test failures. To reduce false positives, the model was instructed to consider only macros relevant to the target translation unit and declared, defined, configured, generated, derived, or documented by the project. Thus, arbitrary compiler-predefined or system-predefined macros were not treated as independent variability dimensions unless the project explicitly exposed them as configurable. The prompt also allowed reporting realistic system/API portability errors, such as missing platform-specific constants, types, declarations, structure fields, or headers, but only when the use was reachable and the project did not provide a fallback or wrapper.

For each target file, the prompt required a JSON object containing a list of detected problems, each with the affected file, minimal failing product or portability condition, active preprocessor condition stack, error description, and supporting evidence. If no failing configuration was found, the model had to return an empty problem list. The complete prompt used in all executions, with \texttt{<path/to/file.c>} replaced by each target file, is available in our replication package~\cite{artefatos}.

\begin{tcolorbox}[
  breakable,
  colback=gray!5,
  colframe=black!60,
  boxrule=0.5pt,
  arc=2pt,
  left=6pt,right=6pt,top=6pt,bottom=6pt
]
\footnotesize
\noindent You are an expert C engineer specialized in configurable and portable C systems implemented with C preprocessor variability (\#if/\#ifdef/\#ifndef/\#elif feature flags) and system API portability. \\

\noindent Project: the C project currently opened in this VS Code workspace. \\

\noindent TARGET FILE \\
\noindent Analyze this target file: \\
\noindent TARGET\_FILE = "\texttt{<path/to/file.c>}" \\

\noindent You may inspect any project file needed to understand this target file, including headers, configuration files, generated config templates, Makefiles, Kconfig files, CMake files, Meson files, portability headers, and other source files. However, report only compilation problems that affect the target file or its translation unit. \\

\noindent LANGUAGE STANDARD \\
\noindent Assume the target file must compile as ISO C99. Use C99 rules for syntax, declarations, types, preprocessing, and semantics when reasoning about compilation errors. \\

\noindent GOAL \\
\noindent Find compilation errors in the target file caused by either: \\
\noindent 1. project-controlled preprocessor variability; or \\
\noindent 2. realistic system/API portability variability, where a system-provided macro, constant, type, function declaration, struct field, or header may be unavailable and the project does not provide a fallback. \\
\noindent A variability-induced compilation error is a compilation failure that occurs only for some assignment of relevant project-controlled macros or realistic platform/API availability conditions, while at least one other assignment avoids the failure. \\

\noindent MACRO AND API SCOPE \\
\noindent Analyze macros that appear in \#if/\#ifdef/\#ifndef/\#elif directives in the target file, headers included by the target file, configuration files that affect the target file, and build files that define configuration macros for the target file. \\
\noindent Treat a macro as a project-controlled variability dimension only if it appears in the project and is declared, defined, configured, generated, derived, or documented by the project itself. \\
\noindent Do not introduce completely hypothetical macros unrelated to the target translation unit. \\
\noindent Do not treat arbitrary compiler-predefined or system-predefined macros as independent variability dimensions unless the project explicitly declares, defines, derives, wraps, or documents them as configurable within the project. \\
\noindent Also analyze system/library API constants, macros, declarations, struct fields, and headers used directly by the target translation unit when their absence is a realistic portability boundary. \\
\noindent \ldots \\
\noindent FALSE POSITIVE REDUCTION RULES \\
\noindent Before reporting a problem, check whether the reported code is reachable in the target file's translation unit, the macro or API availability condition is relevant to the target file, the failure is a C99 compilation error and not only a warning, the failure is not link-time-only, and the relevant declaration and use are truly compiled together in the target file's translation unit. \\
\noindent If the evidence is weak, do not report the issue as a problem. \\
\noindent IMPORTANT RULES \\
\noindent Do not edit files. \\
\noindent Do not propose patches. \\
\noindent Do not report style issues, warnings, runtime bugs, memory bugs, link-time errors, or test failures. \\
\noindent Focus only on compilation errors under a standard C99 compiler such as gcc or clang with -std=c99. \\
\noindent Do not analyze the whole project broadly. Use other files only as context for the target file. \\
\noindent \ldots \\
\noindent OUTPUT FORMAT \\
\noindent Return a simple JSON object with exactly this key: \\
\noindent \{ \\
\noindent "problems": [ \\
\noindent \{ \\
\noindent "file": "target file or included file where the problem occurs", \\
\noindent "failing\_product": "minimal boolean assignment over project-controlled macros or realistic portability condition", \\
\noindent "condition\_stack": "active \#if/\#elif/\#else/\#endif path or unguarded system/API use that reaches the failing code", \\
\noindent "error": "short description of the C99 compilation error", \\
\noindent "evidence": "compiler command and diagnostic, or short source-code evidence if compilation was not run" \\
\noindent \} \\
\noindent ] \\
\noindent \} \\
\noindent If no failing configuration is found, return: \\
\noindent \{ \\
\noindent "problems": [] \\
\noindent \} \\
\noindent Do not include markdown. \\
\noindent Do not include any keys other than problems. \\
\noindent Inside each problem, do not include any keys other than file, failing\_product, condition\_stack, error, and evidence. \\
\end{tcolorbox}
\label{lst:codex-prompt-summary}

\subsubsection{Model and Execution Setting}
\label{sec:models-real}

We evaluated \codex{} as a foundation-model assistant integrated into the development environment. The goal was to approximate a workflow in which a developer asks an IDE-based assistant to inspect one target file within the context of an opened project workspace. We executed \codex{} through the OpenAI Codex extension in  Visual Studio Code 1.124.0 (Universal) on macOS Tahoe 26.2, using medium reasoning effort and retaining the remaining interface settings at their defaults. The analysis was conducted in June~2026. The machine was a Mac Mini with an M4 Pro processor and 64\,GB of unified memory. The model was accessed through an OpenAI Plus account. 

For each subject, we cloned the corresponding repository and checked out the pre-fix project revision recorded in the dataset. The target C file was opened in Visual Studio Code and analyzed using the same prompt template. The model could inspect relevant workspace context, including headers, configuration files, build files, portability layers, and other source files, but it was instructed to report only compilation problems affecting the target file or its translation unit. We did not provide the model with the commit identifier, bug identifier, mutation category, mutation description, compiler diagnostic, expected answer, or location of the injected edit.

\subsubsection{Evaluation Procedure}
\label{sec:evaluation-procedure-real}

We evaluated each response against the target-fault oracle established during dataset construction. For the five historical subjects, the oracle is based on the documented bug and its corresponding correction. For the 95 mutation-based subjects, it records the deliberately modified program element, the intended compilation-fault pattern, and the relevant source and conditional-compilation context. This oracle does not represent exhaustive compilation of all project configurations.

For each subject, \codex{} returned a JSON object containing a \texttt{problems} array. A subject was counted as detected only when at least one reported problem was aligned with the target fault. Alignment required the response to identify the affected file or translation unit and describe a compilation problem consistent with the historical bug or deliberately introduced mutation. For local syntax mutations, the response had to identify the corresponding syntax-level compilation problem. For faults involving renamed identifiers, inconsistent declarations and uses, function interfaces, invalid member access, or feature-dependent visibility, it had to relate the reported failure to the relevant program elements or conditional context.

An empty \texttt{problems} array was counted as a miss. Reports unrelated to the target fault and generic observations about conditional compilation that did not identify the relevant problem were also counted as misses. The wording of the model did not need to reproduce the oracle description or compiler diagnostic verbatim, provided that the reported root cause was consistent with the target fault. One author manually assessed target-fault alignment using these criteria. A second author reviewed all classifications. This procedure involves judgment and is discussed as a construct-validity threat. The authors discussed the disagreements and reached consensus; a third author was
available for adjudication when necessary.

The target-fault oracle includes preprocessing, parsing, syntax, declaration, type-checking, and name-binding problems expected to manifest as C99 compilation failures in the target file or translation unit. We exclude warnings, style issues, runtime and memory defects, link-time errors, test failures, and build-system failures. Study~II therefore evaluates localized target-fault detection rather than whole-system build correctness.

Because all 100 subjects contain a target fault, the reported detection rate corresponds to recall over faulty subjects. Study~II does not include fault-free controls and therefore cannot measure whether the model would also report unrelated or nonexistent problems in correct files. We consequently do not report precision, specificity, false-positive rate, F$_1$-score, or overall classification accuracy.

Study~II does not evaluate repair. Although some responses informally suggested possible changes, the prompt instructed the model not to edit files or propose patches. Without reproducible builds and representative project configurations, we could not reliably distinguish an incorrect repair from a failure caused by missing dependencies, generated headers, platform assumptions, or incomplete build environments. We therefore leave whole-system repair validation for future work.

\subsection{RQ$_{3}$. Target-File-Centered Detection in Real Configurable Projects}
\label{sec:results-real}

Overall, \codex{} produced an aligned detection for 94 of the 100 Study~II subjects. It detected four of the five historical bugs and 90 of the 95 mutation-based subjects. Because all subjects contain a target fault and no fault-free controls are included, the reported 94.0\% is a subject-level detection rate, not binary-classification accuracy or an estimate of deployment performance.
Among the six missed subjects, only one response returned an empty \texttt{problems} array. In the other five cases, the model reported one or more other potential compilation problems, but none was aligned with the target fault. 

Across the 100 subjects, \codex{} returned 141 candidate
problems, with a median of one, a mean of 1.41, and a
maximum of five problems per subject. In 68 of the 94
detected subjects, the response contained exactly one
problem, and that problem was aligned with the target fault.
In the remaining 26 detected subjects, the target-aligned
report appeared together with 37 additional reports. Among
the six missed subjects, five responses contained ten reports
in total, but none was aligned with the target fault; only one
response returned an empty problem list. This distribution
indicates that the observed target-fault detection rate does
not appear to result from indiscriminately returning large
lists of candidate problems. However, the additional reports
were not independently validated against complete project
builds and are therefore treated as unvalidated reports
requiring developer triage, rather than as measured true or
false positives.

\subsubsection{Results by Fault Category}
\label{s:study2-results-category}

Table~\ref{tab:study2-results-category} reports the descriptive detection results by fault category. The model detected all evaluated subjects in eight mutation categories: function interface/call, identifier/name-binding, declaration/type-specifier, expression/operator, lexical/punctuation, control-structure, initialization/literal, and preprocessor/feature-condition. Several of these categories contain relatively few subjects, so their 100\% rates should not be interpreted as stable estimates of category-level performance. The six misses occurred in four groups: three pointer/member/array-access mutations, one delimiter/block-structure mutation, one return-value/control-transfer mutation, and one historical bug. Pointer/member/array access had the largest absolute number of misses, with 16 of 19 subjects detected. The historical subset contained only five subjects, of which four were detected; its 80.0\% rate is therefore descriptive and should not be treated as a stable estimate of performance on naturally occurring bugs.

\subsubsection{Results by Project}
\label{s:study2-results-system}

Table~\ref{tab:study2-results-system} reports the descriptive results by project. \codex{} detected all 20 target faults in BusyBox and OpenSSL, 19 of 20 in Vim, 18 of 20 in Apache HTTP Server, and 17 of 20 in Gnuplot. These differences may reflect a combination of file size, fault category, project-specific code structure, and variability density; the study was not designed to isolate a causal project effect.

\begin{table}[t]
\centering
\caption{Study~II detection results by fault provenance and configurable project.}
\label{tab:study2-results-system}
\scriptsize
\setlength{\tabcolsep}{4pt}
\renewcommand{\arraystretch}{0.95}
\begin{tabular}{lrrr}
\toprule
\textbf{} & \textbf{Total} & \textbf{Detected} & \textbf{Rate} \\
\midrule
Historical bugs & 5 & 4 & 80.0\% \\
Mutation-based subjects & 95 & 90 & 94.7\% \\
\midrule
BusyBox & 20 & 20 & 100.0\% \\
Gnuplot & 20 & 17 & 85.0\% \\
Apache HTTP Server & 20 & 18 & 90.0\% \\
OpenSSL & 20 & 20 & 100.0\% \\
Vim & 20 & 19 & 95.0\% \\
\midrule
\textbf{Overall} & \textbf{100} & \textbf{94} & \textbf{94.0\%} \\
\bottomrule
\end{tabular}
\end{table}

\subsubsection{Results by Subject Characteristics}
\label{s:study2-results-characteristics}

Table~\ref{tab:study2-results-characteristics} presents an exploratory stratification by file-level and target-line characteristics. LOC, distinct macros, and conditional-compilation annotations are grouped by quartiles. Maximum nesting depth and active macros at the affected line are grouped according to their observed distributions.

\begin{table}[t]
\centering
\caption{Detection results by subject characteristics. LOC, macros, and
conditional annotations are grouped by quartiles.}
\label{tab:study2-results-characteristics}

\scriptsize
\setlength{\tabcolsep}{3pt}
\renewcommand{\arraystretch}{0.95}

\begin{tabular}{@{}llrrr@{}}
\toprule
\textbf{Characteristic} &
\textbf{Group} &
\textbf{\# Subjects} &
\textbf{Detected} &
\textbf{Rate} \\
\midrule

\multirow{4}{*}{LOC}
  & 42--311     & 25 & 24 & 96.0\% \\
  & 312--647    & 25 & 24 & 96.0\% \\
  & 648--1,793  & 25 & 24 & 96.0\% \\
  & 1,794--7,238 & 25 & 22 & 88.0\% \\
\midrule

\multirow{4}{*}{Distinct macros}
  & 1--2   & 29 & 28 & 96.6\% \\
  & 3--5   & 23 & 23 & 100.0\% \\
  & 6--13  & 24 & 24 & 100.0\% \\
  & 14--50 & 24 & 19 & 79.2\% \\
\midrule

\multirow{4}{*}{Cond. annotations}
  & 1--3    & 26 & 25 & 96.2\% \\
  & 4--9    & 26 & 26 & 100.0\% \\
  & 10--26  & 23 & 23 & 100.0\% \\
  & 27--230 & 25 & 20 & 80.0\% \\
\midrule

\multirow{2}{*}{Max. nesting depth}
  & $\leq 2$ & 76 & 73 & 96.1\% \\
  & $> 2$    & 24 & 21 & 87.5\% \\
\midrule

\multirow{5}{*}{Active macros at fault}
  & 0 & 1  & 1  & 100.0\% \\
  & 1 & 82 & 78 & 95.1\% \\
  & 2 & 12 & 11 & 91.7\% \\
  & 3 & 4  & 3  & 75.0\% \\
  & 4 & 1  & 1  & 100.0\% \\
\bottomrule
\end{tabular}
\end{table}

Detection remained above 95\% in most lower and middle groups, but was lower in the largest LOC quartile, the highest macro-count quartile, the highest conditional-annotation quartile, and the group with nesting depth greater than two. The groups based on active macros at the affected line were highly unbalanced and therefore do not support meaningful comparisons.
Overall, lower descriptive rates appear in the largest-file, highest-macro, highest-annotation, and deeper-nesting groups. However, the benchmark contains only six misses, and characteristics such as LOC, macro count, project, nesting, and fault category are correlated. The observed differences should therefore be interpreted as exploratory associations rather than evidence that any individual characteristic caused the missed detections.
Under the evaluated target-file-centered IDE-based setting with repository context, \codex{} identified a problem aligned with the target fault in 94 of 100 subjects: four of five historical bugs and 90 of 95 mutation-based subjects. These results provide promising evidence for localized detection in files from real configurable projects. They do not establish false-positive behavior, whole-system build correctness, exhaustive configuration-space analysis, or performance on a large population of naturally occurring bugs.

\subsection{Discussion}
\label{sec:discussion-real}

\subsubsection{Selected Detection Cases}
\label{s:study2-discussion-qualitative}

We examine three selected subjects to illustrate distinct detection outcomes and forms of interaction between target faults and conditional-compilation structure. Subject~59 comes from Apache HTTP Server. The affected
file\footnote{\href{https://github.com/apache/httpd/blob/93667b2bf98a0ee851eb37f774bd857a1ef9b1ff/modules/aaa/mod_auth_dbm.c}{Apache HTTP Server Project: File \texttt{mod\_auth\_dbm.c} at commit \texttt{93667b2}}}
contains alternative DBM implementations selected by
\texttt{AP\_AUTH\_DBM\_USE\_APR}. The mutation adds an extra opening
brace inside this conditional region, leaving the surrounding function
structurally unbalanced and causing later function definitions to be
parsed as nested within \texttt{get\_dbm\_pw}. \codex{} identified this
target-aligned problem as a missing closing brace in the APR branch and
reported the resulting downstream compiler diagnostics. A nearby
developer comment---``sorry for the obscurity''---acknowledges the
non-obvious control-flow arrangement in the original code. The case
therefore illustrates how even a non-nested, disciplined conditional
region can make a simple delimiter mutation harder to interpret, although
the comment alone does not establish code obscurity as the cause of the
fault or of the model's reasoning.

The only missed historical bug occurred in Gnuplot's \texttt{gplt\_x11.c} (Subject~21), one of the largest and most variability-dense
files in the benchmark, with 7,238 LOC, 50 distinct conditional-compilation macros, and 190 conditional annotations. Its maximum preprocessor nesting depth is four, and six annotations occur in structurally undisciplined regions. The target fault itself, however, lies in a disciplined, depth-one
\texttt{\#ifndef USE\_X11\_MULTIBYTE} region near the end of the file. It is a typographical inconsistency in which \texttt{disp} is used instead of the declared variable \texttt{dpy}, producing an undeclared-identifier error. \codex{} reported three other plausible configuration-specific compilation problems involving \texttt{USE\_MOUSE}, \texttt{CRIPPLED\_SELECT}, and \texttt{VMS}, all at locations unrelated to the historical fault. The case therefore does not suggest that deep local nesting was required to identify the target. Rather, it illustrates that, in a large file containing many conditional regions and plausible diagnostics, the assistant may reason about several feature-dependent problems while still failing to prioritize or
identify the target fault. Because this is a single subject, we do not interpret file size, variability density, or report competition as the cause of the miss.

Subject~95 comes from Vim's
\texttt{clipboard.c}.\footnote{\href{https://github.com/vim/vim/blob/679c2c01faab2b2106ecd7c1038efa031158b93b/src/clipboard.c}{Vim Project: File \texttt{clipboard.c} at commit \texttt{679c2c0}}} The 4,072-LOC file contains 20 distinct conditional-compilation macros and 106 conditional annotations, with a maximum nesting depth of four. The target mutation removes an opening brace from a disciplined depth-two region controlled by \texttt{FEAT\_CLIPBOARD} and by the alternative features \texttt{FEAT\_XCLIPBOARD} and \texttt{FEAT\_WAYLAND\_CLIPBOARD}. When either clipboard backend is enabled, the unmatched closing brace terminates
\texttt{clip\_gen\_own\_selection} prematurely and leaves a subsequent \texttt{return} statement outside any function. In the original open-ended analysis, \codex{} returned no problem and therefore missed the target fault. In a later diagnostic interaction, we explicitly named
\texttt{clip\_gen\_own\_selection} and asked whether it was missing an opening brace. The model then correctly explained the delimiter imbalance and attributed its earlier miss to examining a configuration in which the affected feature-dependent branch was disabled. This follow-up was not counted in Study~II because the prompt supplied both the suspicious function and the defect type, and the model's retrospective explanation does not establish its actual reasoning process. The case nevertheless illustrates the distinction between independent fault discovery and explaining a fault after its location and nature have been provided: the model could explain the interaction among the conditional branches once directed to it, but did not discover it during the original analysis.

\subsubsection{Scope and Limitations}
\label{s:study2-discussion-effectiveness}

Study~II suggests that an IDE-based assistant with repository context can provide useful target-file-centered detection support in real configurable projects. Unlike the compact snippets of Study~I, these subjects contain real declarations, includes, portability mechanisms, conditional-compilation structures, and unrelated surrounding code. The detection of faults involving identifiers, declarations, types, and function interfaces is consistent with reasoning about relationships among program elements beyond isolated tokens in the evaluated subjects.
Detection rates were lower in groups containing larger files, more distinct macros and conditional annotations, and greater nesting depth. This pattern is consistent with the hypothesis that variability-dense files impose a greater search and reasoning burden, but Study~II was not designed to estimate the independent effects of these characteristics. Only six subjects were missed, and file characteristics are correlated with project and fault category; we therefore treat these associations as hypotheses for future controlled studies rather than causal findings.

The results should also be interpreted in light of the benchmark composition. Ninety-five of the 100 subjects contain controlled mutations, and the five historical bugs are too few to establish performance on naturally occurring faults by category. The remaining misses, including one historical bug, reinforce that \codex{} should complement rather than replace compiler-based or symbolic variability-aware analyses. The findings support an assistive target-file-centered workflow with repository context, not claims of complete fault detection in configurable projects.
Because Study~II contains no fault-free controls, it estimates target-fault detection sensitivity rather than precision or false-positive behavior. Additional reported problems were not independently validated, so report volume cannot substitute for a false-positive analysis. Study~II also evaluates detection only; reliable repair assessment would require reproducible complete-project builds, feasible configurations, generated artifacts, dependencies, platform assumptions, and regression tests.

We also explored whether the \typechef{} comparison could be extended to Study~II. In a preliminary attempt to analyze a BusyBox source file, we supplied the target file, the project headers, and the include directories that we could identify from the project structure. \typechef{} began preprocessing the file and traversed several BusyBox and system headers. However, although it emitted a generic message indicating that parsing had succeeded, the execution log showed that the lexer had failed for all configurations, and the complete source file was therefore not effectively parsed or analyzed. We were unable to reconstruct and configure the project-specific build and analysis environment sufficiently for \typechef{} to complete a valid analysis. We therefore do not report this exploratory attempt as an empirical result and restrict the empirical comparison to Study~I, where the setup is controlled and reproducible and all configurations can be validated exhaustively using a compiler-based oracle.

\subsubsection{Implications}
\label{s:study2-implications}

For developers, foundation models may support early detection and triage when inspecting feature-dependent code with repository context. Their reports should nevertheless be treated as hypotheses requiring compiler-based or symbolic validation, particularly because Study~II does not estimate false-positive behavior or whole-system build correctness.
For tool developers, the findings motivate hybrid workflows in which compilers and variability-aware tools provide systematic configuration evidence and validation, while foundation models help explain, prioritize, and localize findings and propose candidate changes. Structured output can facilitate such integration, but syntactic JSON validity does not establish that a reported problem is correct or aligned with the relevant fault.
For researchers, future evaluations should combine the control of Study~I with the realism of Study~II through larger collections of historical bugs, matched fault-free controls, reproducible project builds, feasible feature configurations, systematic configuration sampling, and whole-system repair validation. Synthetic benchmarks should also be audited for lexical and structural properties that predict oracle labels, preferably using counterfactual tests with edit-matched controls before model comparisons.

\section{Threats to Validity}
\label{s:threats-validity}

Several threats to the validity of our study must be considered~\cite{threats-llms-icse-nier-2024}.

\subsection{Construct validity}

\revision{Study~I evaluates two related constructs. Subject-level detection asks whether at least one configuration fails, whereas configuration-level coverage compares the reported and compiler-derived failing sets. The binary task is affected by an unintended benchmark property: every compilable subject contains a conditional \texttt{\#define}, whereas no non-compiling subject does, allowing a trivial rule to separate the labels perfectly. This does not invalidate the compiler-derived oracle, but it prevents binary accuracy and recall from establishing variability-aware reasoning on their own. The paired counterfactual audit in Section~\ref{sec:counterfactual} found prediction-change rates comparable to sham edits and test--retest variation and therefore did not detect a material effect of this property on \gptoss{} in the evaluated sample. However, the audit covers one model, one sample, and one property and cannot exclude smaller effects, reliance outside the sample, or dependence on other correlated structures.}

\revision{Configuration-level coverage introduces a separate measurement threat. The prompt requests the minimal set of failing products the model can confidently justify, and responses use either complete assignments or partial assignments that function as presence conditions. Scoring only complete assignments would measure compliance with an enumeration format and discard informative partial reports. We therefore expand each partial assignment to the set of concrete products it denotes and report both interpretations in Section~\ref{s:results-rq1}. This assumes that unconstrained macros may take either value; if the model intended a narrower set, the expansion overstates recall and understates precision. The observed 80.5\% micro-precision for presence-condition responses is consistent with some over-coverage. Neither interpretation estimates performance under a prompt that explicitly requires exhaustive enumeration, which would require a separate experiment.}

\revision{The Study~I restoration oracle establishes only compilability. A subject is restored end-to-end when it contains a non-compiling configuration, the model reports an error, supplies a nonempty candidate, and the candidate compiles under every derived configuration under ISO C99. This criterion does not establish semantic correctness, behavioral equivalence, preservation of intended functionality or variability semantics, or absence of newly introduced non-compilation defects. We therefore use the terms compiler-accepted transformation and compilability restoration rather than correct repair.}

\revision{Study~II uses a target-fault-alignment oracle rather than exhaustive build validation. For the five historical subjects, the oracle follows the documented bug and correction; for the 95 mutation-based subjects, it records the deliberately modified element and intended fault pattern, which were manually inspected in context. A response counts as a detection only when it identifies a problem connected to the target declaration, use, condition, type inconsistency, or feature interaction. Because every subject contains a known target fault and no fault-free controls are included, the reported 94.0\% measures target-fault detection sensitivity, not precision, specificity, false-positive rate, or deployment accuracy. We also did not build every project under all relevant configurations, so Study~II does not establish whole-system build correctness.}

\subsection{Internal validity}

\revision{The results may depend on prompts, model versions, temperatures, APIs, local inference frameworks, IDE behavior, parsing decisions, and evaluation judgments. We mitigate these threats by documenting execution settings and criteria and by releasing prompts, responses, scripts, hashes, and supporting artifacts~\cite{artefatos}. As part of Study~I, a deterministic compiler-based audit records the compiler invocation, treats a configuration as non-compiling only when the process returns a nonzero exit code, records warnings separately, and validates candidate transformations under every derived configuration. This design reduces ambiguity in the oracle and restoration criterion, although defects in implementation or artifact collection cannot be entirely excluded.}

\revision{For the 357-subject sample used in the repeated-run and complementary-model analyses, a source-level audit compiled every Boolean assignment and established that 167 subjects compile under every configuration and 190 contain at least one non-compiling configuration, totaling 1,048 failing configurations. It also identified 146 configurations across 20 subjects that emit diagnostics while returning a zero exit code and are therefore warnings rather than failures. The scripts and outputs are included in the replication package, but this additional source-level audit was limited to the sampled subjects.}

\revision{The repeated-run analysis joins subject identifiers with the deterministic compiler-derived oracle and counts missing or empty candidates as unsuccessful restorations. The counterfactual audit compiles the original, cue, and sham variants and retains only subjects whose exact failing sets remain unchanged. Repeated executions estimate model variability, but five runs cannot fully characterize the output distribution. The counterfactual audit was conducted in July~2026, whereas the full-benchmark and repeated-run evaluations were conducted in January~2026, using the same prompt, temperature, and local model identifier. Although the identifier was unchanged, differences in distributed weights or the inference framework cannot be excluded. In Study~II, mutation inspection and target-alignment classification also involve judgment, especially when a response reports a related symptom rather than the expected root cause; explicit criteria reduce but do not eliminate this subjectivity.}

\revision{Execution conditions are not identical across models. \gptoss{} was run locally at temperature 0.4 through an inference framework that may normalize formatting, whereas \gemini{} was run through a hosted API at temperature 0 with provider-side schema enforcement. These differences affect output validity and may affect predictions. \gemini{} was also evaluated once, so its values have no run-to-run interval; because eight subjects account for all of its restoration failures, its 95.8\% rate should not be interpreted as a more precise estimate than the repeated-run results for \gptoss{}.}

\revision{All Study~I oracle checks use \clang{} with ISO C99 explicitly enabled, and warnings are not treated as failures. Compiler defects, platform-specific behavior, and implementation-specific diagnostics remain possible~\cite{DBLP:conf/pldi/YangCER11}. Results may therefore differ with another compiler, language mode, platform, or diagnostic scope.}

\subsection{External validity}

\revision{Study~I uses small synthetic configurable C snippets that enable exhaustive validation but omit much of the complexity of configurable systems, including cross-file dependencies, build-system interactions, generated headers, feature-model constraints, macro-expansion chains, deep nesting, undisciplined annotations, long-range declaration--use relationships, and semantic coupling among features. No exact duplicates were found under basic normalization, but the 5,000 subjects instantiate only 45 abstract templates, and 382 highly similar pairs involve 120 subjects. Together with the \texttt{\#define} association, these findings show that the benchmark contains repeated and label-correlated structures and should be interpreted as a controlled benchmark rather than as 5,000 independent examples of naturally occurring code.}

\revision{The benchmark was generated by a foundation model from 30 small configurable C examples curated in prior work~\cite{DBLP:conf/sbes/AlbuquerqueG024}. Compiler-based validation provides an independent oracle, and the evaluated models differ from the generator, but the generated distribution may still be more regular or more recognizable to foundation models than naturally occurring bugs. The counterfactual audit reduces concern about one discovered property for \gptoss{}, but it does not eliminate broader LLM-shaped benchmark bias or establish the same result for \gemini{}. Seed-to-instance provenance was not retained and could not be reconstructed reliably, so we cannot report performance per seed; moreover, only 7 of the 15 seed error labels have a plausible counterpart among the normalized first diagnostics in the benchmark.}

\revision{The complementary-model comparison covers 357 of the \csSmall{} subjects, two model families, and a single \gemini{} execution. Reporting style may vary across models, prompts, and decoding settings, so the observed configuration-coverage values characterize the evaluated settings rather than stable model properties. The shared difficulty with warning-only diagnostics should likewise be confirmed with additional models and datasets.}

\revision{Study~II improves realism by evaluating larger files from BusyBox, Gnuplot, Apache HTTP Server, OpenSSL, and Vim with repository context, but it includes only five historical bugs and 95 injected faults from five systems. The injected variants preserve real project files and variability context but are not naturally occurring bugs, and their representativeness depends on the selected mutation patterns~\cite{Just2014MutantsRealFaults,Papadakis2018MutationScores,Petrovic2021MutationTesting}. Study~II does not evaluate complete builds, cross-file fault propagation, feasible configuration spaces, fault-free files, or candidate repairs. Its aggregate result must therefore not be interpreted as performance on 100 naturally occurring bugs or as evidence of whole-system analysis.}

\revision{Neither study reconstructs complete feature models. Study~I treats every Boolean assignment over the snippet macros as valid, which enables exhaustive and reproducible validation but may include infeasible configurations. Study~II likewise does not characterize the complete valid product spaces of the selected projects. Future evaluations should combine foundation models with feature-model extraction, build-system analysis, or variability-aware tools.}

\subsection{Conclusion validity}

\revision{The two studies support conclusions at different scopes. Study~I provides evidence about subject-level detection, configuration reporting, repeated-run stability, and compiler-accepted transformations for small synthetic snippets. Study~II provides evidence about localized target-fault detection in faulty real-project files. Neither study establishes that foundation models can replace compilers or symbolic variability-aware analyses, and Study~II does not validate generated transformations.}

\revision{The distinction between subject-level and configuration-level findings is central. The benchmark labels are trivially predictable from the \texttt{\#define} property even though the counterfactual audit did not detect a material cue effect for \gptoss{} in the evaluated sample. At the configuration level, presence-condition responses cover 99.4\% of the failing configurations in that subgroup, whereas explicit enumeration under a minimal-set instruction remains incomplete and accounts for 93.2\% of the residual false negatives. The supported conclusion is therefore not that models cannot characterize affected configurations, but that complete and verified sets still require external compiler-based or symbolic validation. Likewise, compilability restoration does not imply semantic correctness, and the \typechef{} comparison is descriptive for the documented benchmark and setup rather than a general ranking.}

\revision{The stability analysis uses one fixed sample of 357 subjects. Its \textit{pass@\(k\)} results are retrospective and oracle-dependent: they show that at least one successful response exists among the first \(k\) attempts but do not identify which response should be selected in practice. The \gemini{} evaluation uses the same sample and one execution, so it has no independent stability estimate. Broader samples, additional models and runs, fault-free controls, independent annotations, and complete project builds are needed before stronger conclusions can be drawn.}

\revision{Training-data contamination also remains possible. Different generation and evaluation models reduce direct same-model self-evaluation in Study~I but do not eliminate shared model-family patterns. In Study~II, the model did not receive commit identifiers, diagnostics, mutation descriptions, expected answers, or fault categories, and the injected variants did not previously exist in the repositories. Nevertheless, unchanged project code, historical bugs, and similar fault patterns may have appeared in public training corpora. Finally, model versions, APIs, IDE extensions, and service-side behavior evolve, so the findings characterize the evaluated versions and environments rather than permanent model capabilities.}

\section{Related Work}
\label{s:related}

Several variability-aware tools have been proposed to detect syntactic and type-related errors in configurable C systems, notably \typechef{}~\cite{typechef} and \textsc{SuperC}~\cite{superc}. These tools use static analyses that preserve conditional-compilation information to identify errors that arise only under particular configurations.
\revision{
Their application to real projects may require reconstructing relevant build context, including include paths, macro definitions, platform assumptions, generated headers, and configuration constraints. In contrast, we investigate whether foundation models can support variability-induced compilation-error analysis through prompt- and IDE-based interfaces with less project-specific analysis setup. This does not eliminate maintenance effort: model-based approaches
still require prompt maintenance, model-version tracking, and repeated empirical validation as models, APIs, and deployment environments evolve.

In Section~\ref{s:discussion-typechef}, we compare
\gptoss{} with \typechef{} on the synthetic Study~I
benchmark. The approaches exhibit different failure patterns:
\typechef{} is highly precise under the evaluated
parser-oriented setup, whereas \gptoss{} reports more of the
subjects considered faulty by the broader compiler-based
oracle and can generate human-readable explanations and
candidate transformations. However, the binary labels are
perfectly separable using the presence of
\texttt{\#define}, and the two approaches cover different
diagnostic scopes. We therefore use the comparison to
characterize complementary behavior under the evaluated
setup, not to establish the superiority of foundation models
over variability-aware analyses.
}

Abal et al.~\cite{Abal18,abal-2014} proposed a catalog of 98
variability bugs in configurable C systems and studied their
characteristics, including several classes of compilation errors. This catalog has since supported empirical studies of
variability-induced faults in real-world systems.
\revision{
Building on this line of work, Study~I generates \csSmall{} synthetic configurable C snippets, comprising \csSmallCE{} faulty and \csSmallOK{} compiling subjects. These snippets provide a controlled benchmark in which all macro assignments can be compiled exhaustively. However, because the generation process may transform, simplify, or combine the original fault patterns, we do not treat the resulting benchmark as a collection of naturally occurring bugs.
Study~II complements this benchmark with files from BusyBox, Gnuplot,
Apache HTTP Server, OpenSSL, and Vim. For each project, we selected the
revision immediately preceding a commit that fixed a historical
variability-related compilation bug. The original faulty file was
included together with 19 additional files from the same revision
modified using fault patterns inspired by real variability bugs.
Thus, Study~II combines five historical faults with 95 controlled
mutation-based subjects in realistic project-file contexts.
}

Medeiros et al.~\cite{flavio-gpce-2013} proposed a technique to identify syntactic errors in configurable C systems. Subsequently, Medeiros et al.~\cite{flavio-gpce2015} introduced a static analysis for detecting undeclared-variable uses under conditional compilation. Both approaches identified real compilation errors in configurable C
systems. Braz et al.~\cite{Braz:2016,Braz:2018} later proposed a change-aware analysis for detecting compilation errors introduced during software evolution.
\revision{
These studies rely on purpose-built static analyses and provide systematic reasoning about conditional-compilation structures. Our work investigates a complementary approach based on foundation models. Study~I evaluates detection and compiler-accepted repair on controlled synthetic snippets, whereas Study~II evaluates localized detection in files from real configurable projects. The evaluated models can provide
explanations, localization hints, and candidate repairs, but they do not offer the soundness, completeness, or exhaustive configuration-space coverage of symbolic analyses.
}

The repair component of our work is related to the broader field of Automated Program Repair~\cite{apr-cacm2019,apr-survey2018}. Recent studies have explored LLMs for repairing programming errors, including compilation failures. A recent survey provides a systematic overview of LLM-enabled compilation, introduces a taxonomy of existing approaches, and discusses their potential to broaden compiler
development~\cite{compiler-llm-hpc2026}.
\revision{
Our work contributes to this emerging area by focusing specifically on compilation errors induced by feature variability in configurable C code. Repair is evaluated only in Study~I, where all configurations of the synthetic snippets can be compiled. Study~II evaluates detection only, because validating repairs in real configurable projects would
require reproducible builds, project-specific dependencies, generated configuration headers, platform assumptions, representative feature configurations, and regression tests.
}

A closely related work is RustAssistant~\cite{rust-llm-icse2025}, which combines compiler diagnostics with iterative LLM interactions to repair compilation errors in Rust programs. Its evaluation includes microbenchmarks, Stack Overflow examples, and GitHub commits from popular Rust crates, and shows that compiler feedback can guide LLMs
toward compilable patches.
\revision{
Our work differs from RustAssistant in both the target problem and evaluation setting. RustAssistant repairs compilation failures in concrete Rust programs, whereas we study feature-dependent compilation errors in configurable C code. These errors require reasoning about conditional-compilation regions, feature interactions, and declarations whose availability changes across configurations. We also compare \gptoss{} with the variability-aware parser \typechef{} in Study~I and evaluate \codex{} for target-fault detection in real configurable projects through a target-file-centered IDE workflow that can inspect relevant repository context beyond the analyzed file.
}

Another closely related study is Albuquerque et al.~\cite{DBLP:conf/sbes/AlbuquerqueG024}, who manually evaluated 30 configurable examples to investigate whether four LLMs could identify compilation errors. \revision{Our work builds directly on that study by using its 30 configurable examples as prompting seeds for the construction of the Study~I benchmark. We extend the initial investigation in three directions. First, Study~I constructs and exhaustively compiler-validates \csSmall{} synthetic configurable C snippets and evaluates \gptoss{}, with complementary analyses involving \typechef{} and a stratified sample of \gemini{}. Second, it evaluates whether model-generated transformations restore compilability across all derived configurations in this controlled setting. Third, Study~II evaluates \codex{} on 100 faulty file-level subjects from five real configurable projects, comprising five historical bugs and 95 mutation-based subjects, through a target-file-centered IDE workflow with access to relevant repository context. Nevertheless, Study~II remains a per-file target-fault detection evaluation rather than a whole-system, configuration-aware build analysis.}

\section{Conclusions}
\label{s:conclusions}

\revision{This article investigated foundation models for detecting variability-induced compilation errors and, in a controlled setting, restoring compilability. Study~I evaluated \csSmall{} synthetic configurable C snippets with an exhaustive compiler-based oracle, whereas Study~II evaluated localized target-fault detection in 100 file-level subjects from five mature configurable C projects. Under a prompt requesting only a minimal confidently justified set, explicit enumerations covered 29.5\% of the affected configurations, whereas presence-condition reports covered 99.4\%. \gptoss{} restored compilability for 1,930 of 2,665 faulty snippets (72.4\%), and, in a stratified sample of 357 subjects, \gemini{} restored 182 of the 190 faulty subjects (95.8\%). In Study~II, \codex{} met the target-fault alignment criterion in 94.0\% of subjects. These results support localized detection and compilability restoration with external validation rather than semantic correctness or exhaustive configuration coverage.}

\revision{Model-generated benchmarks and their scoring pipelines should be audited for provenance, duplication, structural regularities, label-correlated properties, and interpretive assumptions, because such artifacts can make aggregate metrics misleading. The conditional \texttt{\#define} property perfectly separates the binary labels; although the paired counterfactual audit did not detect a material effect on \gptoss{} in the evaluated sample, other correlated structures and broader synthetic-artifact bias remain possible. Moreover, the overlap with \typechef{} reveals complementary failure patterns, supporting hybrid workflows rather than replacement of symbolic analyses. Study~II shows that coding assistants can support localized inspection, explanation, and triage in realistic project files with repository context, including historical and previously nonexistent injected faults. However, evaluations containing only faulty subjects measure target-fault detection sensitivity rather than precision or specificity. Reliable deployment therefore requires fault-free controls, configuration-aware validation, and complete project builds.}

\revision{Overall, foundation models are best positioned as complements to compilers and symbolic variability-aware tools. Models can inspect context, explain likely causes, prioritize suspicious regions, and propose candidate transformations, while deterministic analyses provide systematic configuration evidence and validation. Future work should evaluate hybrid workflows combining foundation models with feature-model and build-system analysis, using broader historical-bug collections, fault-free controls, cross-file scenarios, and configuration-aware tests.}

\section*{Acknowledgments}
% We want to thank the anonymous reviewers for their insightful suggestions. 
%This work was partially supported by CNPq, FAPESQ-PB, and Embrapii grants.
This work was partially supported by INES.IA (National Institute of Science and Technology for Software Engineering Based on and for Artificial Intelligence, \url{www.ines.org.br}) under CNPq grant 408817/2024-0, by CNPq under grants 408040/2025-4 and 403719/2024-0, and by FAPESQ-PB (grant 268/2025). Additional support was provided by the project ``iSOP Base: Investigação e desenvolvimento de base arquitetural e tecnológica da Intelligent Sensing Operating Platform (SOP)'', funded by the CENTRO DE COMPETÊNCIA EMBRAPII VIRTUS EM HARDWARE INTELIGENTE PARA INDÚSTRIA (VIRTUS-CC), with financial resources from the PPI HardwareBR program of MCTI (grant 055/2023), in partnership with EMBRAPII.

\bibliographystyle{IEEEtran}
% Generated by IEEEtran.bst, version: 1.14 (2015/08/26)

\end{document}